\documentclass[%
 reprint,
 amsmath,amssymb,
 aps
floatfix,
]{revtex4-1}

\usepackage{color}
\usepackage{graphicx}% Include figure files
\usepackage{dcolumn}% Align table columns on decimal point
\usepackage{bm}% bold math
\usepackage{subfigure}
\usepackage{caption}
\usepackage{natbib}
\usepackage[bookmarks=false]{hyperref}
\begin{document}

\preprint{APS/123-QED}

\title{Stability of Implanted Transition Metal Dopants in Rock-salt Oxides
}% Force line breaks with \\
%\thanks{A footnote to the article title}%

\author{Debolina Misra}
 \email{mm18ipf03@iitm.ac.in}
\author{Satyesh K. Yadav}%
 \email{satyesh@iitm.ac.in}
\affiliation{%
Department of Metallurgical and Materials Engineering, Indian Institute of Technology Madras, Chennai, 600036, India}%

\date{\today}% It is always \today, today,
             %  but any date may be explicitly specified

\begin{abstract}
Transition metals (TMs) implanted in oxides with rock-salt crystal structures (for example MgO and BaO) are assumed to substitute cations (Mg in case of MgO) from the lattice sites. We show that not all implanted TMs substitute cations but can be stable in interstitial sites as well. Stability of TM (Sc--Zn) dopants in various charge states in MgO and BaO has been investigated in the framework of density functional theory. We propose an effective way to calculate stability of implanted metals that let us predict site preference (interstitial or substitution) of the dopant in the host. We find that two factors govern the preference for an interstitial site: i) relative ionic radius and ii) relative oxygen affinity of cation and the TM dopants. If the radius of the cation is much larger than TM dopant, as in BaO, TM atoms always sit at interstitial sites. On the other hand, if the radius of the cation is comparable to that of the dopant TM, as in case of MgO, the transition of the preferred defect site, from substituting lattice Mg atom (Sc to Mn) to occupying interstitial site (Fe to Zn) is observed. This transition can be attributed to the change in the oxygen affinity of the TM atoms from Sc to Zn. Our results also explain experiments on Ni and Fe atoms implanted in MgO. This is the first-time we have shown that TM dopants can be stable at interstitial sites in stable compounds, which could potentially give rise to exotic properties. 

\end{abstract}

\pacs{Valid PACS appear here}% PACS, the Physics and Astronomy
                             % Classification Scheme.
%\keywords{Suggested keywords}%Use showkeys class option if keyword
                              %display desired
\maketitle

%\tableofcontents

\section{\label{sec:intro}Introduction}
Dopants in semiconductors cause significant changes in their electronic and optical properties, as required for their industrial applications. Among various ways available, ion-beam implantation is a very reliable and popular technique to incorporate impurities in a host lattice \cite{WHITE1989, MARKEVICH2006, PEREZ1984} as it provides controllable selective area doping in the target materials \cite{zno-ionbeam}. The most widely used implanted defects in semiconductors include P, B, As \cite{Ruffell_2005,AMBERIADIS1990651,Nakamura2012,Fujijap,Fujiapl} and Si$^+$ \cite{si_si,si_sio2_positron} implanted thermal oxide films on crystalline Si for photoluminiscence, modified refractive index and optical waveguides, O and N in SiC \cite{Pomaska_2016}, P \cite{p_sio2}, B \cite{b_si}, Si \cite{si_sio2, si+_sio2, guha_1998} and Er \cite{elliman_2002, er_sio2} in SiO$_2$,  Au \cite{au_sio2}, Cu \cite{cu_sio2} in SiO$_2$, Pd in Si \cite{sood_2006} and As in GaAs \cite{as_gaas} for tuning optical and physical properties. 

Doped rock-salt oxides (for example MgO) have been studied extensively \cite{scanlon2007, leejid2014, SCHUT1999, chen-mgo, llusar-mgo, Uberuaga2013, Uberuagaprl2004, shinjap2008, wu-mgo, Schefflerprl, gilbertprb2007, Henkelmanprb2005} for their applications in optical and magnetic sensors, switching devices and as dilute magnetic semiconductors \cite{Rama_2007, Maoz_2011, AZZAZA2014}. Apart from various chemical routes available for creating defects in MgO \cite{AZZAZA2014, mart_2010, mishra_2013, choi_1998}, there have been instances where MgO is implanted with different ions. These include, Au for modifying refractive index of MgO and creating quantum antidots \cite{wang_2003, xu_prl_2002, xu_1999, ZIMMERMAN1998}, Cu and Ni for modifying optical properties \cite{ZIMMERMAN1998, XIANG2006}, He, Ar, Fe, Cr for photoluminiscence \cite{crawford_1973, Skvortsova_2010}, Ne, Ar, Zr, Ru, Si, Cr and Fe for enhancing secondary electron emission yield \cite{Averback-mgo, lee_2014,lee_2016}. MgO implanted with various magnetic impurities like Fe, Cr, Ni, Co have also been reported to exhibit giant magneto resistance (GMR), super-paramagnetism \cite{perez_1983, ZHU2006, Narayan_1981, hayashi_pssa} and  ferromagnetic ordering \cite{Narayan_2008, sharma_2011}. 
\\
To our surprise, studies on ion-implanted rock-salt oxides assume that the dopants occupy only the substitutional sites replacing host lattice cations without any exploration of the other possible sites including interstitials. Possibility of dopant occupying interstitial sites has also been ignored in cases like Li intercalation in rock-slat-structured entropy stabilized oxide \cite{hahn_2018}, citing Pauling's rule \cite{pauling}. Although there has not been any previous study on the site preference of implanted transition metals in rock-salt oxides, one of our recent works \cite{syacta2015} hinted at the possibility of Fe atoms occupying interstitial position in MgO. Here we carry out a systematic investigation of the thermodynamic stability of implanted transition metals (TMs) in MgO. Our work explores the (i) stability of dopants in MgO in both neutral and charged states, (ii) preferred defect sites in the host lattice, and (iii) relation between stability, local distortion and changes in the electronic structure of the host lattice caused by the dopants, for Sc, Ti, V, Cr, Mn, Fe, Co, Ni, Cu and Zn. Besides MgO, we have also studied TM dopants in BaO which has similar crystal structure but larger lattice parameter than MgO. The reason for choosing two oxides is to perform a comparative study in order to understand the effect of interstitial volume on the stability of TM defects. Here we have used density functional theory (DFT) which has long been used efficiently for studying the stability of charged and neutral defects in oxides \cite{VandeWallejap2004, janottiprb2007} by calculating defect formation energies \cite{Freysoldtprb2016,lanyprb2008, ivadyprb2013,lanyprb2008}.
 \\ 
This paper is organized as follows. Section II describes the state-of-the-art computational methodology to calculate stability of TM dopants in semiconductors. In Section III, we report and discuss the results we have obtained from our DFT calculations and finally we conclude in section IV.

\section{\label{sec:method}Methodology}
\subsection{Formation energy of implanted defects}
\begin{figure}[h]
\includegraphics[width=2.9in]{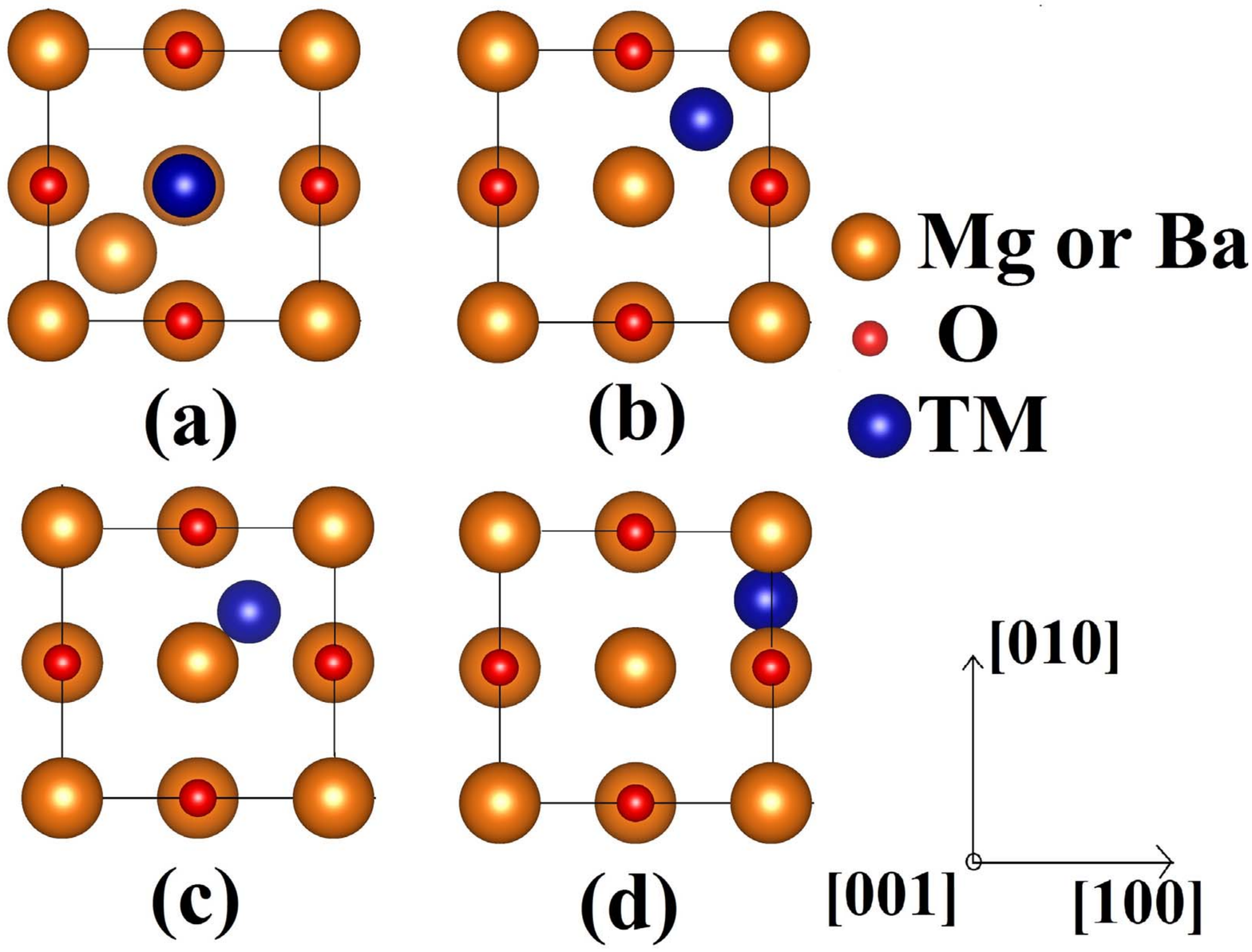}
\caption{Atomic structure of TM dopants in MgO or BaO: (a) TM replaces host cation and the host cation sits in the tetrahedral void; TM dopant sits at (b) middle (c) corner and (d) edge of the tetrahedral void.}
{\label{structure}}
\end{figure}

The focus of this present study is to explore the possibility of implanted TM being stable at the interstitial sites or substituting the host cations from the lattice sites. To do so, we calculate the dopant formation energies for two different atomic configurations: (1) TM placed at host cation site (wyckoff 4a(0.5,0,0.5)) and host atom is pushed to the center of the tetrahedra formed by oxygen atoms as shown in Fig.~\ref{structure}a, (here onwards we would refer this configuration as replacement) and (2) TM at the center of the tetrahedra formed by oxygen atoms (wyckoff 8c(0.25,0.25,0.25)) as shown in Fig.~\ref{structure}b (we would refer this configuration as interstitial site). 
If the first configuration is more stable, then the implanted TM atom replaces the host cation and push them into the interstitial site. The host cation (Mg and Ba in our case) now in interstitial site, can then migrate out of the matrix thus forming substitutional doping. Fast migration of Mg interstitials in MgO upon irradiation has already been reported earlier with a barrier height of 0.32 eV associated with its migration \cite{Blas_prl}. If the second configuration is more stable then TM will not replace host cation and will be stable as an added atom in the interstitial site. It should be noted that often when substitutional formation energy is calculated, substituted atom is removed from the host lattice, this maintains overall stoichiometry of the host lattice. Such approach is a good representation of the system when doping is achieved by mixing compounds so that overall stoichiometry is maintained. But in case of implanting TMs in oxide, there could be more metal in the oxide if TM atom prefers to occupy interstitial site.

The stability of the charged TM dopant in the oxide is assessed by calculating formation energy \cite{Freysoldtprb2016} $E_f^q$ using the following equation \cite{Rampiprl, Freysoldtprb2016, VandeWalleRev2014, BAJAJ2015},
\begin{equation}
E_f^q=E_D^q-E_B-\eta+q(\mu+E_{ref}+{\Delta}V)+E_{corr}^q
\end{equation}
where $E_D^q$ and $E_B$ are the total energies of the defect supercell with charge q and the defect free host supercell, respectively. $\eta$ is the chemical potential of the transition metal atom species. The $'$-$'$ sign before $\eta$ is due to addition of TM in the host. We take both gaseous and crystalline metal energy as chemical potential of transition metal. Choice of gaseous metal as reference is to represent TM ion implanted in host materials. 
$E_{ref}$ is a suitable reference energy which is generally taken to be the valence band maximum (VBM), the energy of the highest occupied level. $\mu$ corresponds to the electronic chemical potential. $\Delta$V is the correction to realign the reference potential in the defect supercell with that in the defect free supercell \cite{VandeWallejap2004}. $E_{corr}^q$ is the correction to the electrostatic interaction and the finite size of the supercell. In this work we have taken only the first-order monopole correction into account.

\subsection{First-principles method}
Formation energy of transition metal dopants, Sc, Ti, V, Cr, Mn, Fe, Co, Ni, Cu and Zn, in charge states 0, +1, +2, +3 and +4 have been calculated using a cubic super cell containing 32 formula units of MgO or BaO. This results in dopant concentration of 3.1\%. Density functional theory (DFT) as implemented in Vienna \textit{Ab initio} Simulation Package (VASP) \cite{kresse1996efficient, kresse1996commat} was used for all calculation, employing  projector-augmented wave (PAW) method \cite{blochl1994projector}. For all cases, spin-polarized calculations were performed. A plane wave cut-off of 500 eV and a \textit{k}-point mesh of 5x5x5 were used for achieving converged results within 10$^{-4}$ eV per atom. All the structures were fully relaxed using the conjugate gradient scheme and relaxations were considered converged when force on each atom was smaller than 0.02 eV/\AA. For calculating the energy of TM atoms in bulk, the most stable structures were considered and sufficient \textit{k} points were taken to reach the convergence. The density of states (DOS) was calculated by the linear tetrahedron method with Bl\"ochl corrections \cite{blochlprb}.

Generalized gradient approximation (GGA) was used to treat the exchange correlation interaction with the Perdew, Burke, and Ernzerhof (PBE) \cite{perdew1996generalized} functional. Although the defect formation energy varies with the choice of functional \cite{Rampiprl}, GGA is known to provide good qualitative results \cite{sharma_prb}. The use of GGA here is justified by the fact that, focus of this paper is restricted to study the general physiochemical trends related to transition metal dopants in stable oxides, and the results should be taken as qualitative. Performing advanced calculations like HSE to get the accurate values of band gaps and defect formation energies are beyond the scope of the present paper. The lattice parameter and the band gap values obtained from our calculation are 4.2 $\AA$ and 4.43 eV for MgO, and 5.62 $\AA$ and 2.12 eV for BaO, which are in good agreement with some earlier predictions performed with the same level of theory \cite{Ertekin_2013,Schleife_2006,tran_2009,AMORIM2006349}. 

\section{Results and Discussion}
\subsection{Site preference}
To explore whether a dopant prefers to replace host cation or sit at interstitial site, we calculate dopant formation energy with the dopant placed at host lattice or in interstitial as shown in Fig.\ref{structure}(a) and Fig.~\ref{structure}(b), respectively. Chemical potentials for the TM atoms in crystalline and gaseous energy references that we used in our calculations, are listed in Table\ref{tab:mu_tm}.
%Table for TM energy references
\begin{table}
  \begin{center}
    \caption{Chemical potentials for TM dopants using crystalline and gaseous energy references}
    \label{tab:mu_tm}
    \begin{tabular}{c|c|c|c|c|c} 
    {\textbf{TM}} & \multicolumn{2}{c|}{\textbf{Energy/atom}} & {\textbf{TM}} & \multicolumn{2}{c}{\textbf{Energy/atom}}\\
    \hline
    & Crystalline & Gaseous & & Crystalline & Gaseous\\
    & (eV) & (eV) & & (eV) & (eV)\\
    \hline
     Sc & -6.199 & -1.776 & Ti & -7.738 & -1.231\\
     V & -8.720 & -0.593 & Cr & -9.118 & -1.300\\
     Mn & -8.907 & -0.707 & Fe & -8.120 & -0.697\\
     Co & -7.033 & -0.688 & Ni & -5.507 & -0.085\\
     Cu & -3.751 & -0.008 & Zn & -1.106 & -0.011\\
     \end{tabular}
  \end{center}
\end{table}

 In MgO we find that while Sc, Ti, V, Cr and Mn prefer to replace lattice Mg atoms and push them into the interstitial sites, Fe, Co, Ni, Cu and Zn prefer to be at the interstitial sites. This preference does not depend on their charge states. However, some TM atoms in their neutral state seem to deviate from the observed trend. Neutral Ni and Mn, which are expected to sit at interstitial site, prefer a corner of the tetrahedral void available in MgO (Fig.\ref{structure}(c)) (wyckoff 32f (0.81,0.688,0.688)). On the other hand, neutral Fe chooses neither the middle nor the corner of the tetrahedra, but prefers to sit in between two lattice oxygen atoms as shown in Fig.\ref{structure}(d)(wyckoff 48g (0.25,0.9,0.25)). This seemingly unusual position is in agreement with previously reported observation employing Mossbauer spectroscopy and density functional theory calculations \cite{Molholt_2014}.

Unlike MgO, no change in the preferred defect site has been observed in BaO for the entire range of TM atoms studied (Sc--Zn). All the stable defects, neutral or charged, prefer to sit at interstitial. However, an interesting trend in the interstitial defect position has been observed as the atomic radius of the defect changes. While TM atoms with relatively larger atomic radii (Sc--Mn) occupy the middle of the tetrahedral void (Fig.~\ref{structure}(b)), Fe--Cu, with relatively smaller atomic radii prefer to sit at the corner of the tetrahedra (Fig.\ref{structure}(c)). 

 \subsection{Stability of TM dopants}
% MgO Formation Energy Plot
\begin{figure}[t!]
    \centering
    \begin{subfigure}
        \centering
        \includegraphics[height=1.3in]{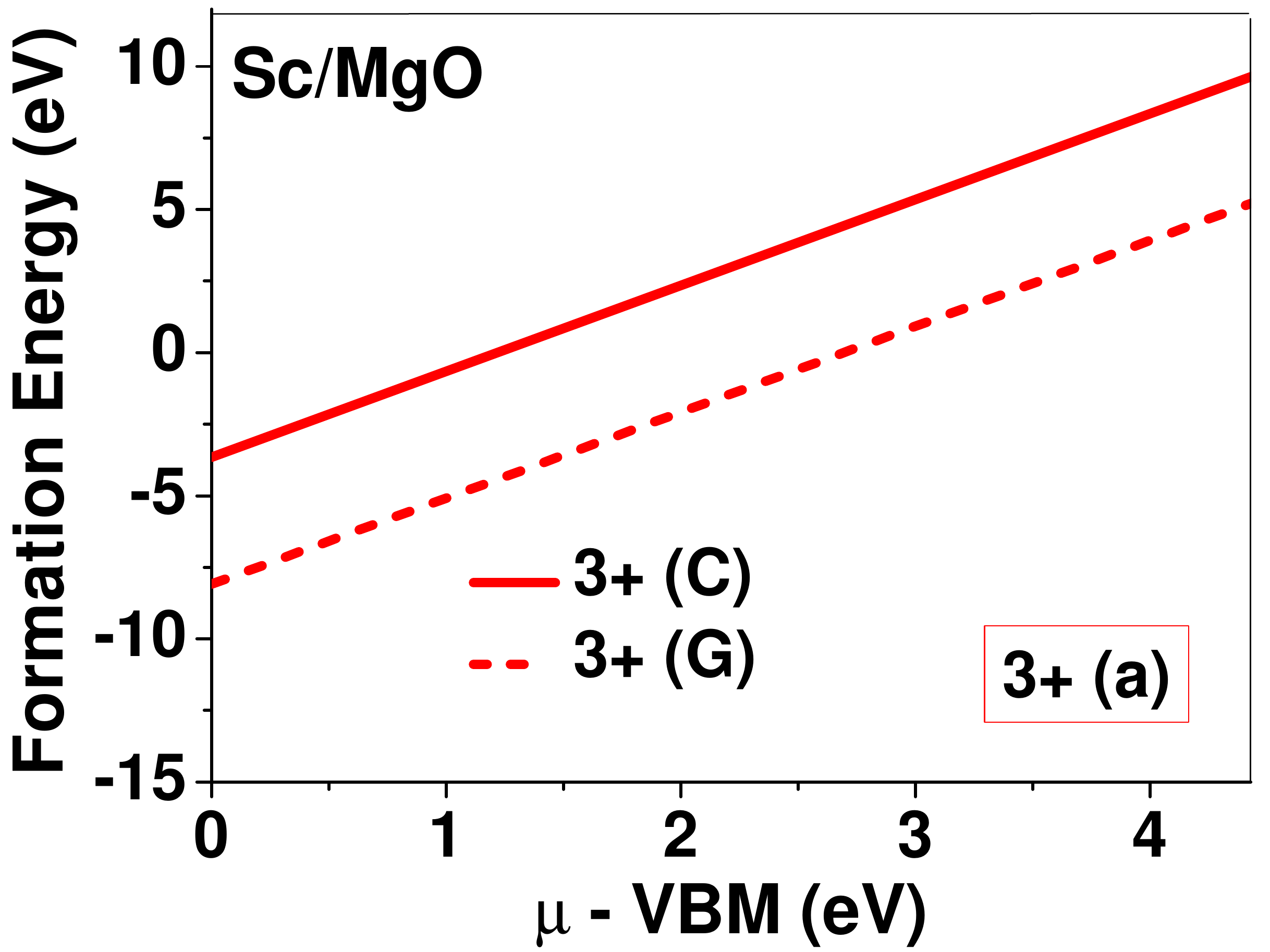}
    \end{subfigure}%
    \hfill
    \begin{subfigure}
        \centering
        \includegraphics[height=1.31in]{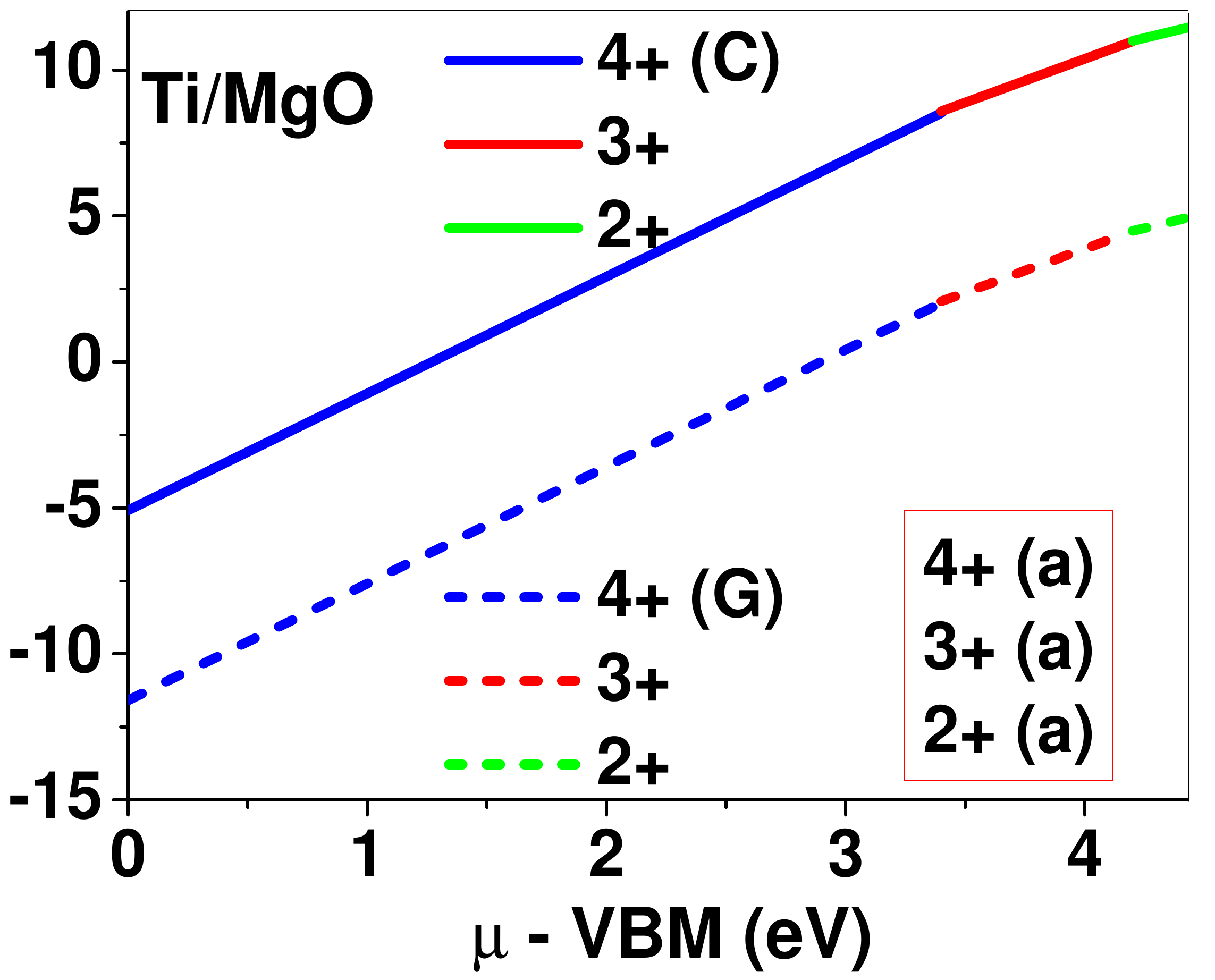}
    \end{subfigure}
    \hfill
    \begin{subfigure}
        \centering
        \includegraphics[height=1.295in]{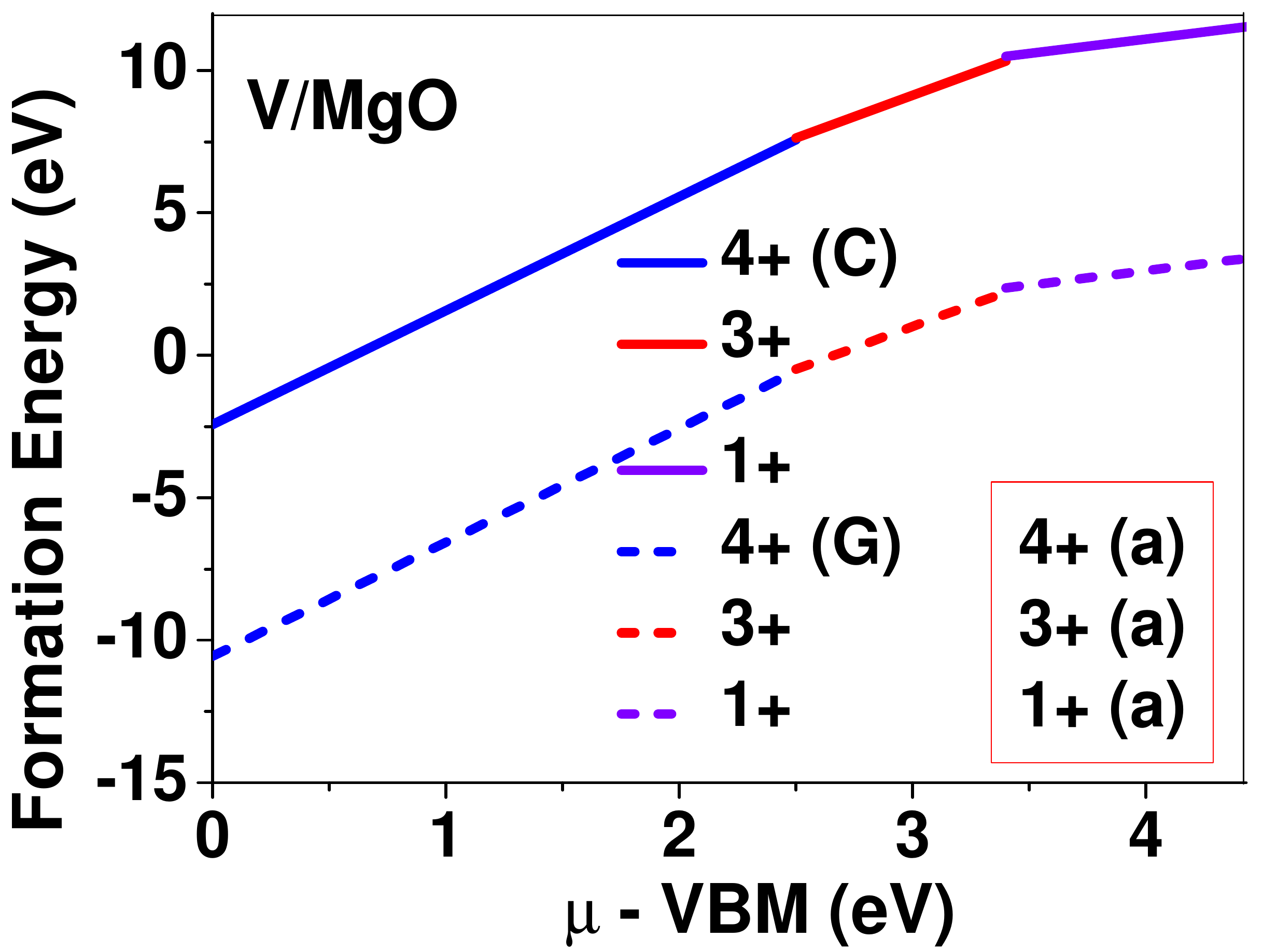}
    \end{subfigure}
    \hfill
    \begin{subfigure}
        \centering
        \includegraphics[height=1.3in]{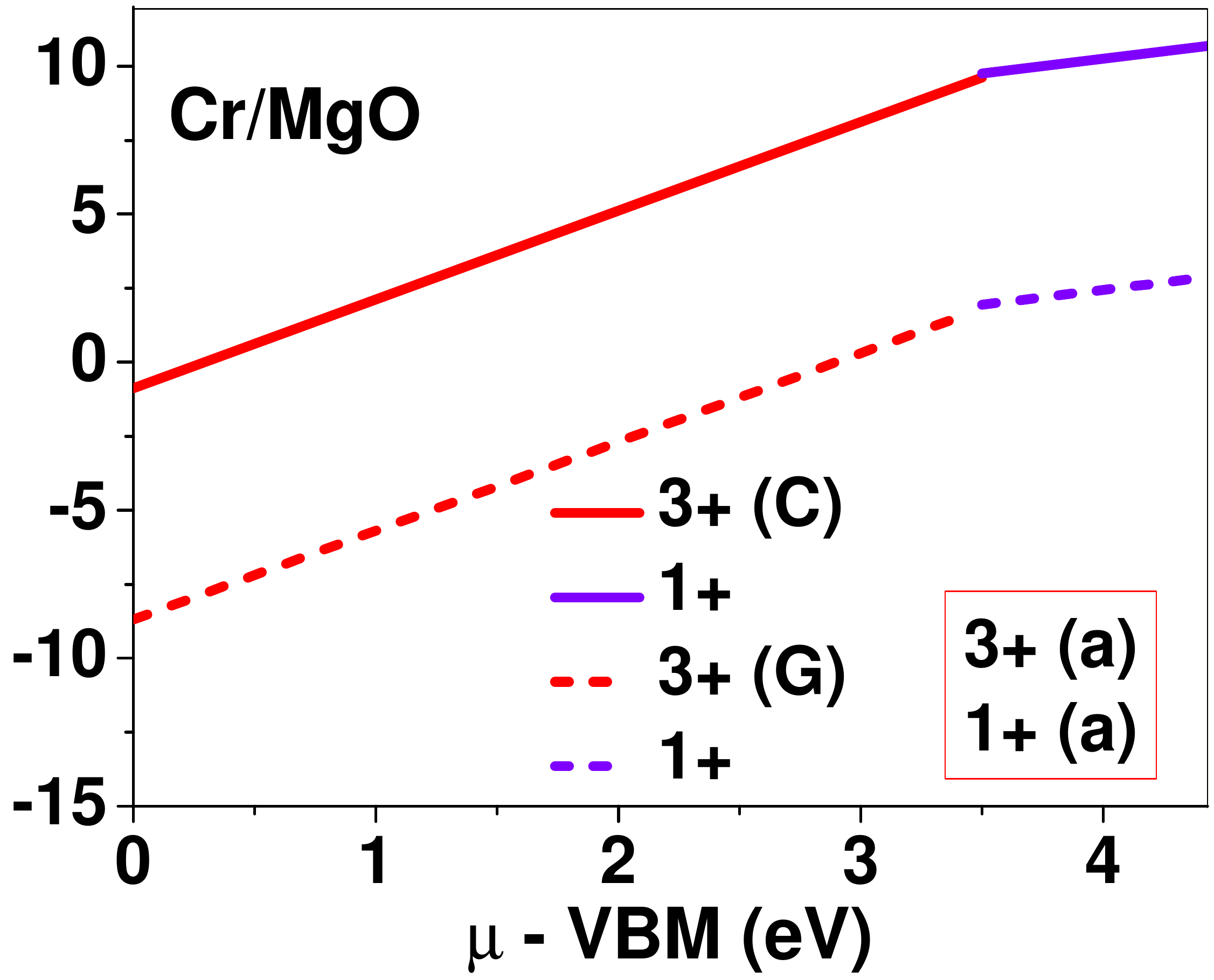}
    \end{subfigure}
    \hfill
    \begin{subfigure}
        \centering
        \includegraphics[height=1.3in]{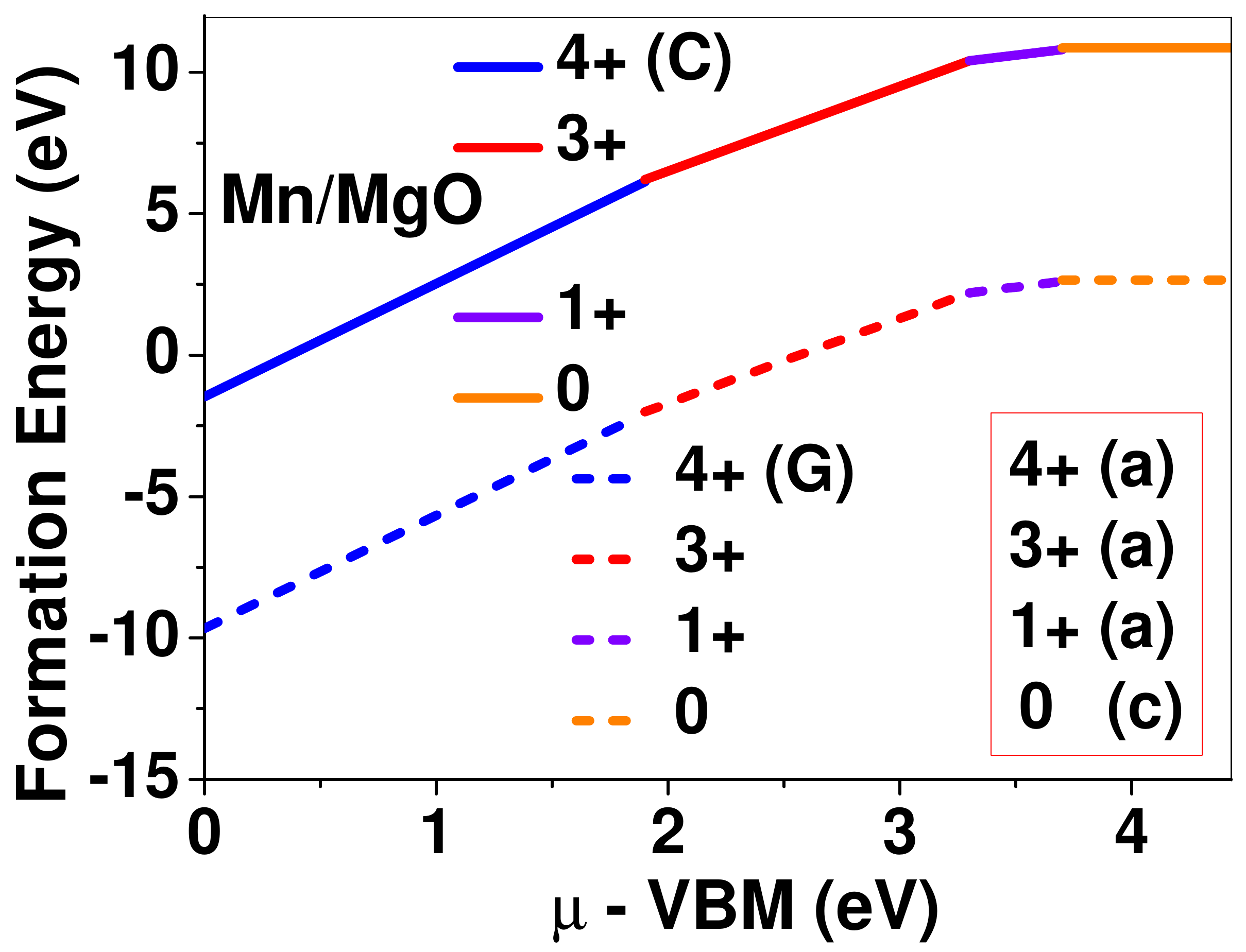}
    \end{subfigure}
    \hfill
    \begin{subfigure}
        \centering
        \includegraphics[height=1.31in]{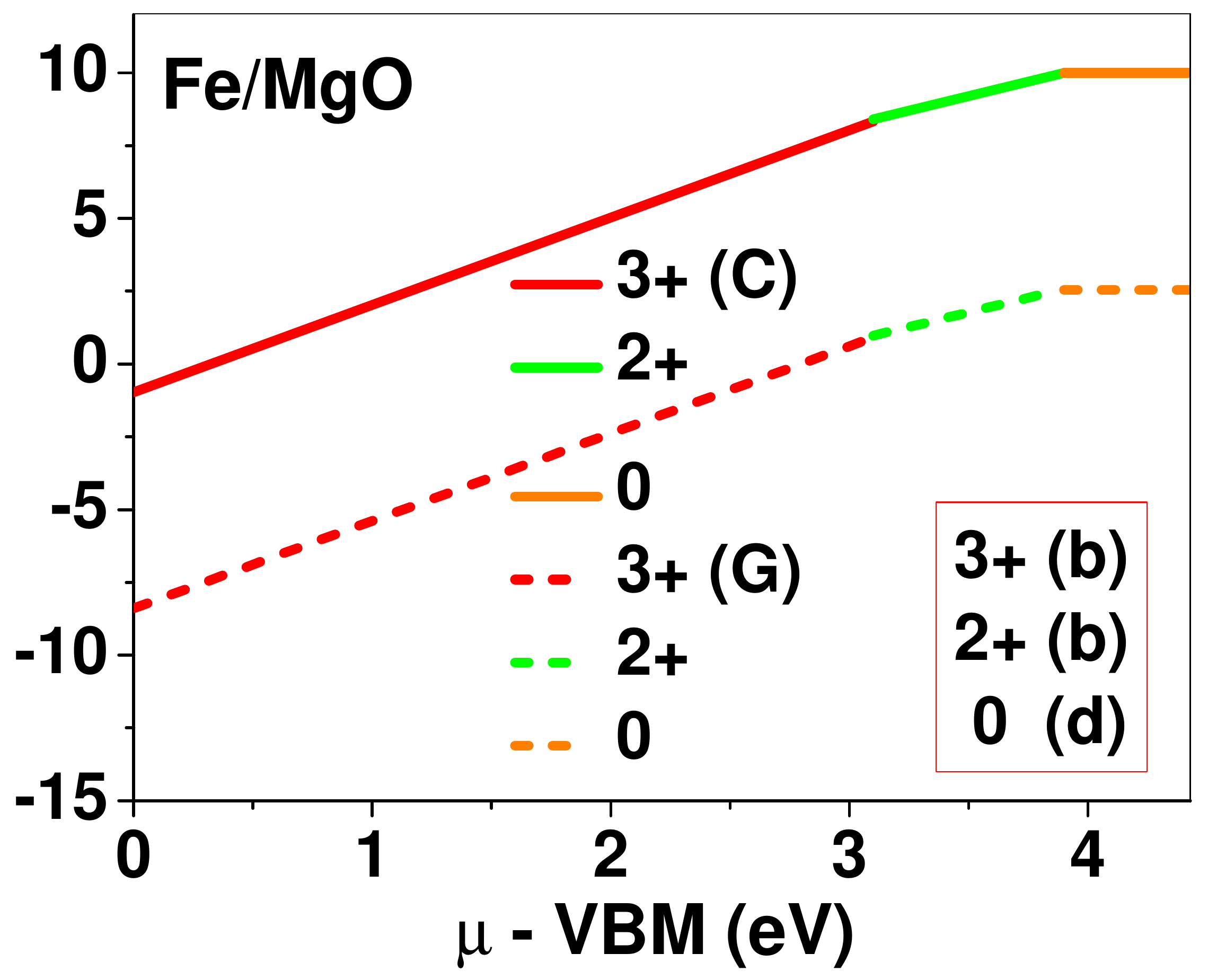}
    \end{subfigure}
    \hfill
    \begin{subfigure}
        \centering
        \includegraphics[height=1.3in]{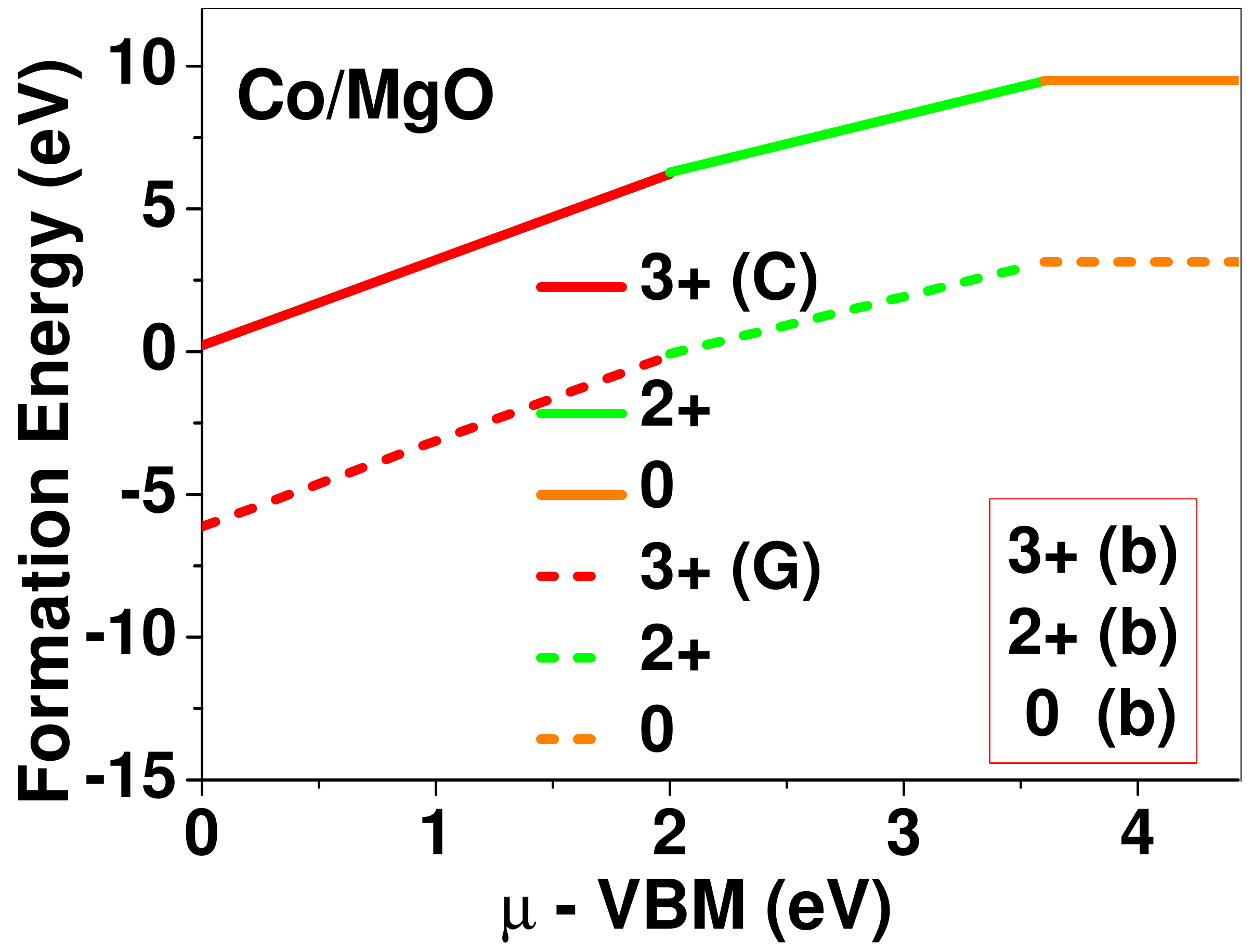}
    \end{subfigure}
    \hfill
    \begin{subfigure}
        \centering
        \includegraphics[height=1.31in]{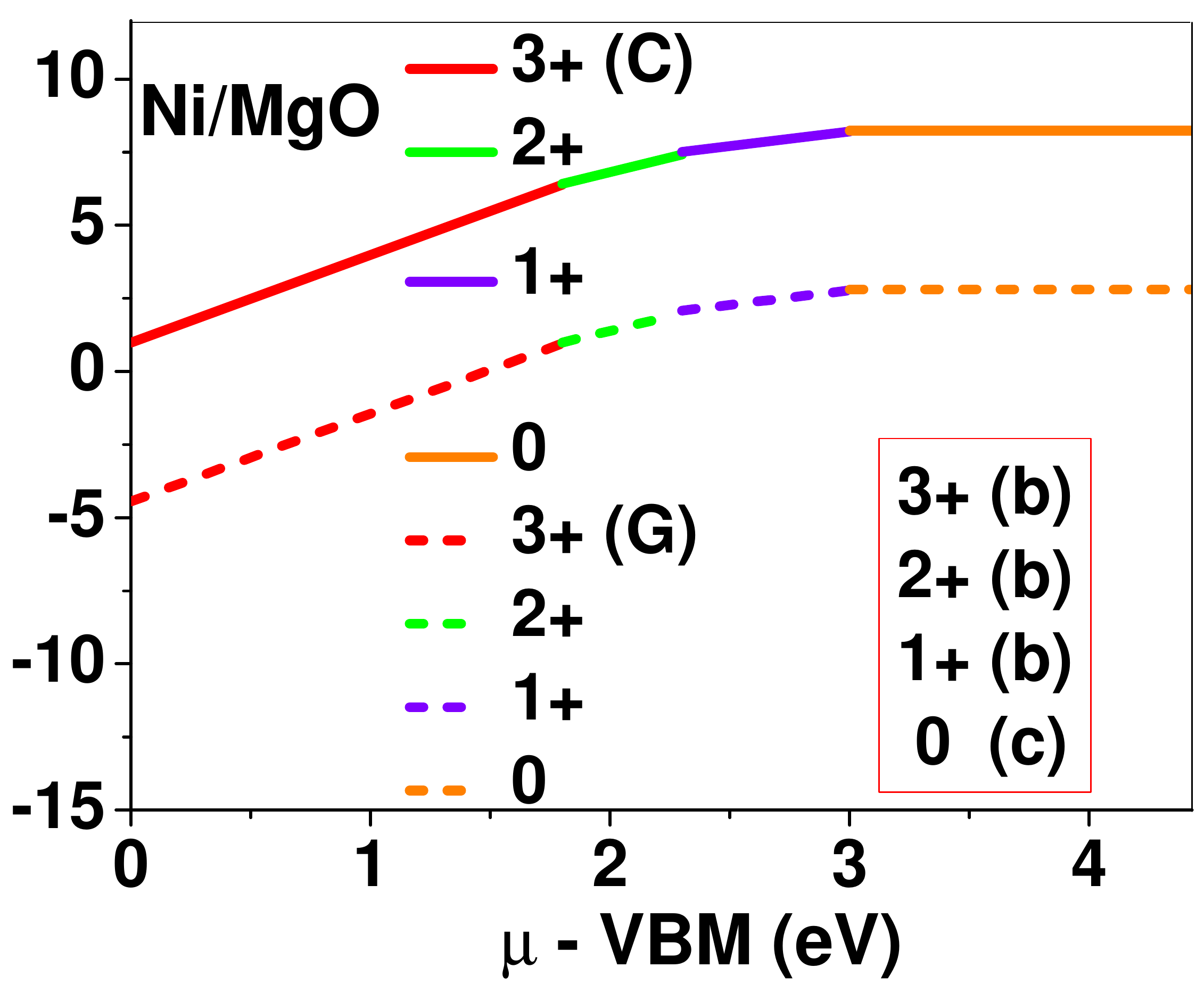}
    \end{subfigure}
    \hfill
    \begin{subfigure}
        \centering
        \includegraphics[height=1.295in]{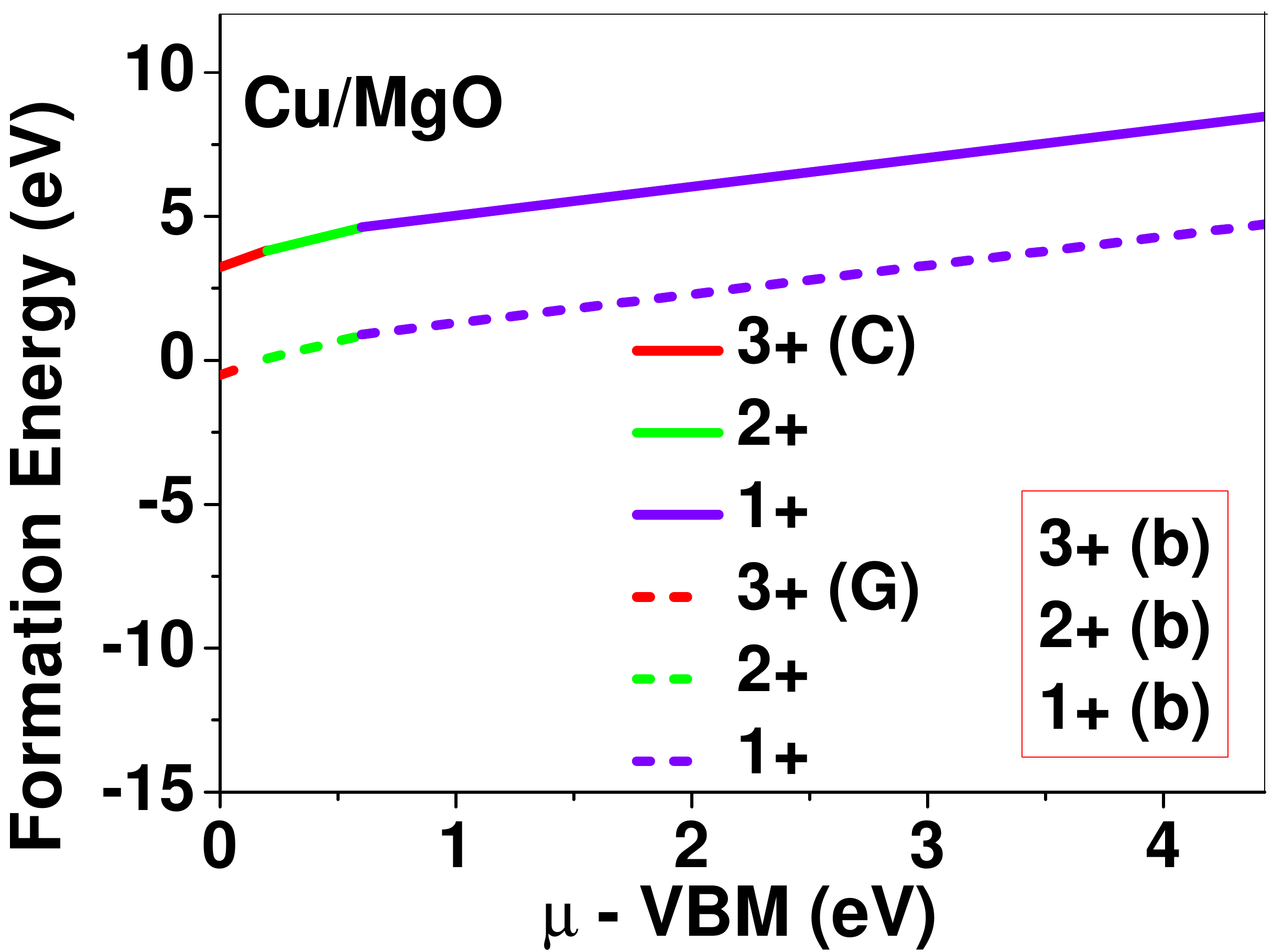}
    \end{subfigure}
    \hfill
    \begin{subfigure}
        \centering
        \includegraphics[height=1.3in]{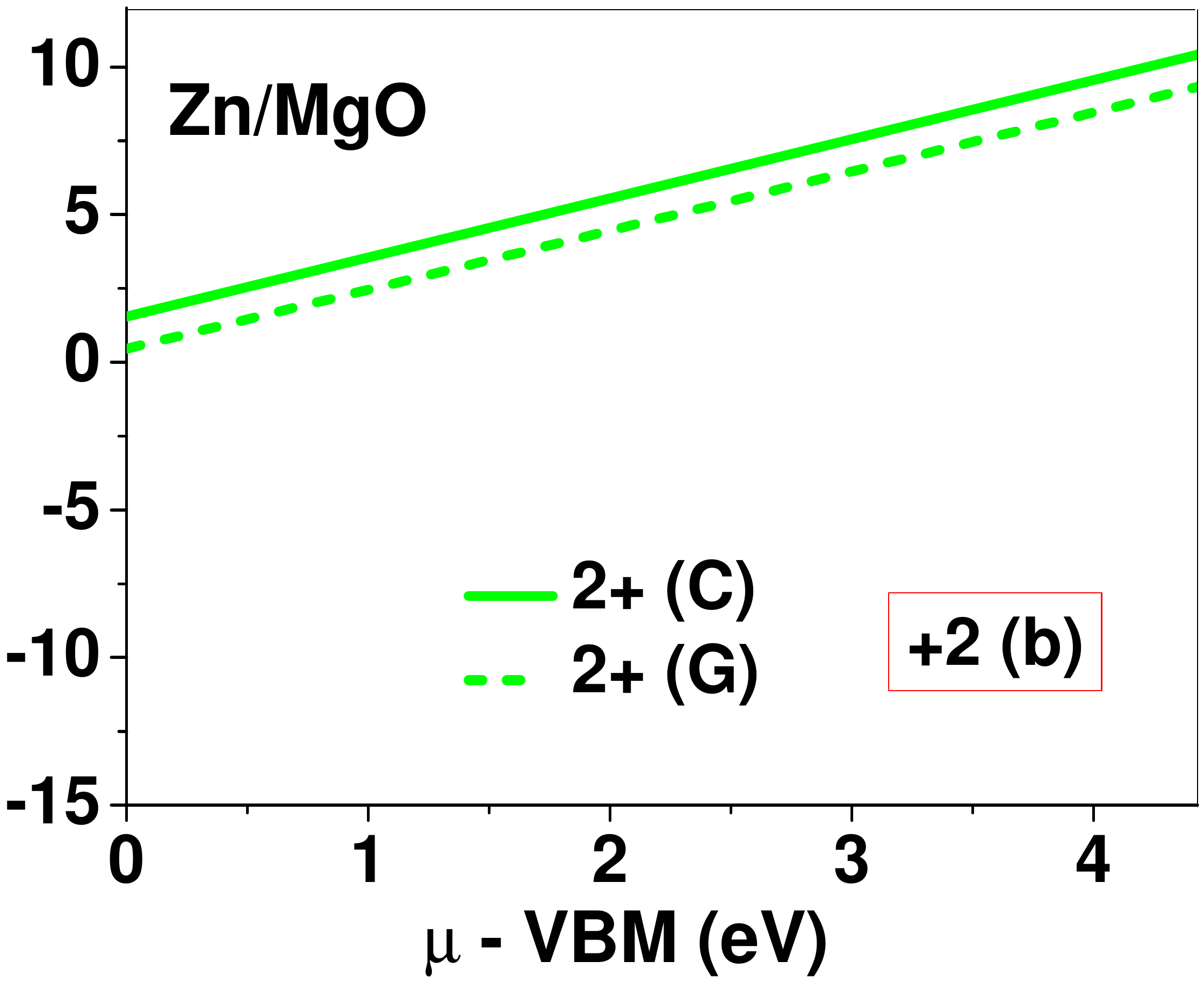}
    \end{subfigure}
    \caption{Formation energy for the neutral and charged dopants in MgO as a function of electronic chemical potential $\mu$; here $\mu$ is referenced to the valence band maximum (VBM). Dashed and solid lines represent the TM dopant formation energies, with gaseous(G) and crystalline(C) metal energy as references, respectively. The preferred defect site for each charge state is indicated in the bracket. Here (a-d) corresponds to Fig.~\ref{structure} (a-d) that shows the atomic structure of TM dopants in rock-salt oxides.}
{\label{mgo_form}}
\end{figure}

%Bao formation energy plot
\begin{figure}[t!]
    \centering
    \begin{subfigure}
        \centering
        \includegraphics[height=1.3in]{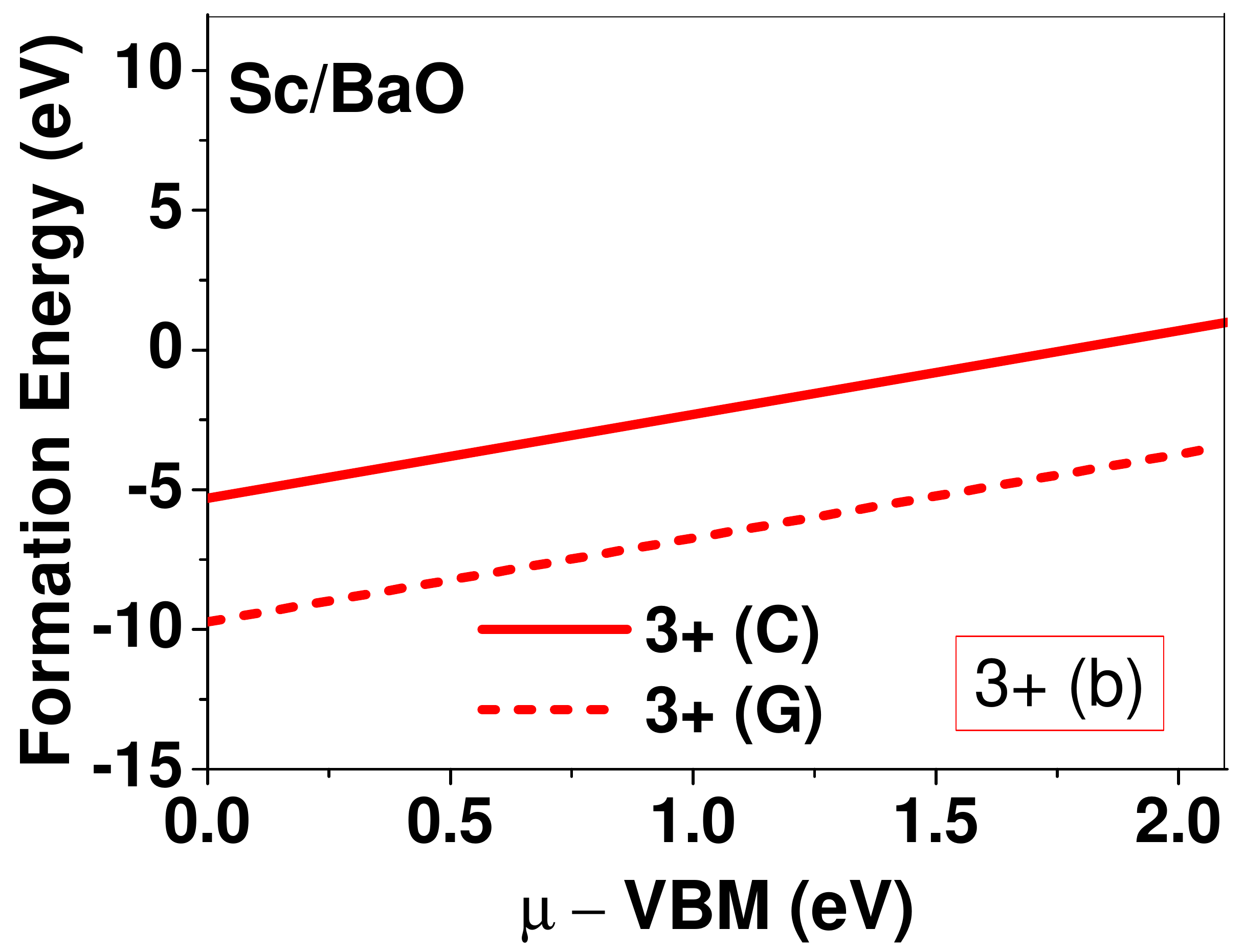}
    \end{subfigure}%
    \hfill
    \begin{subfigure}
        \centering
        \includegraphics[height=1.31in]{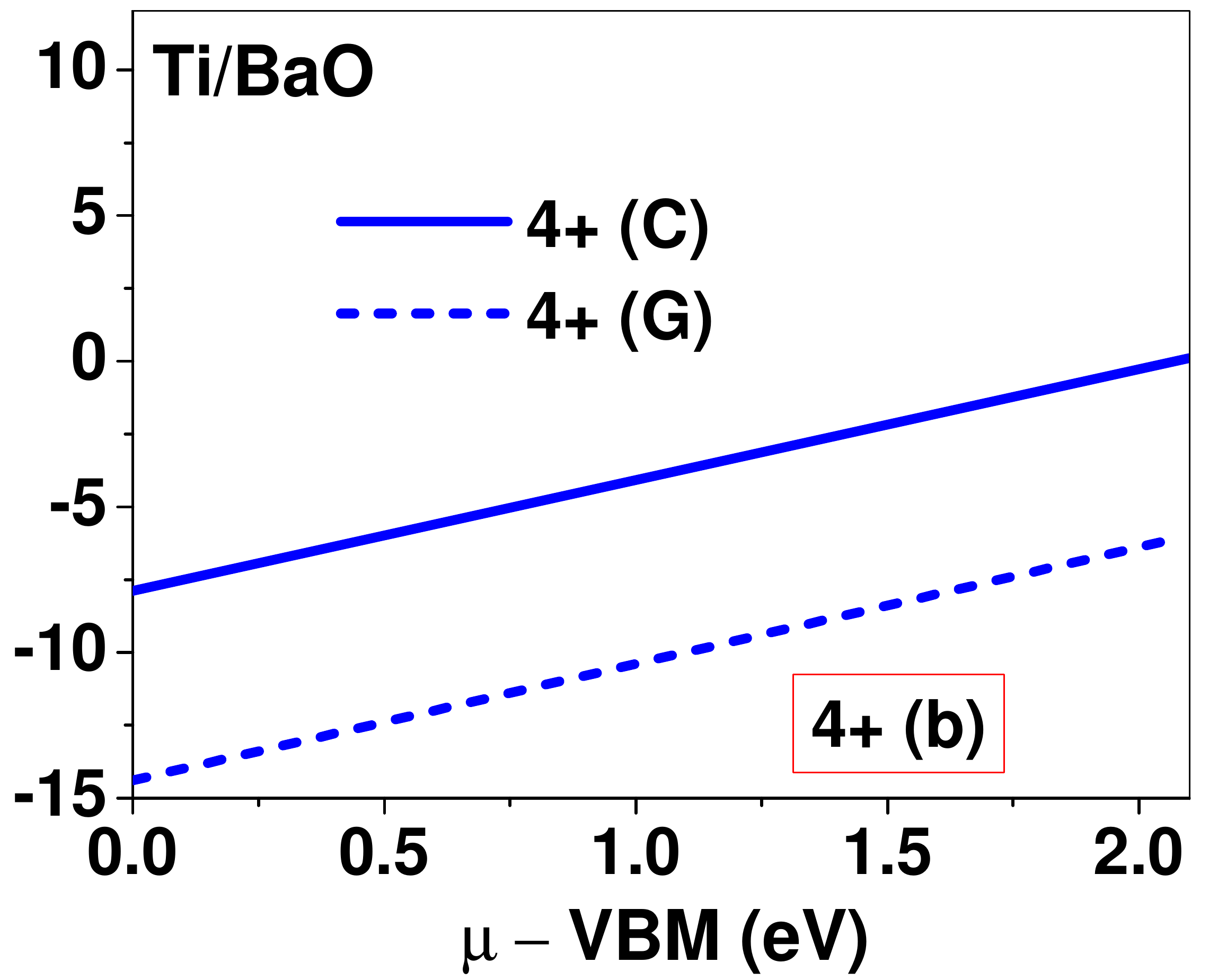}
    \end{subfigure}
    \hfill
    \begin{subfigure}
        \centering
        \includegraphics[height=1.3in]{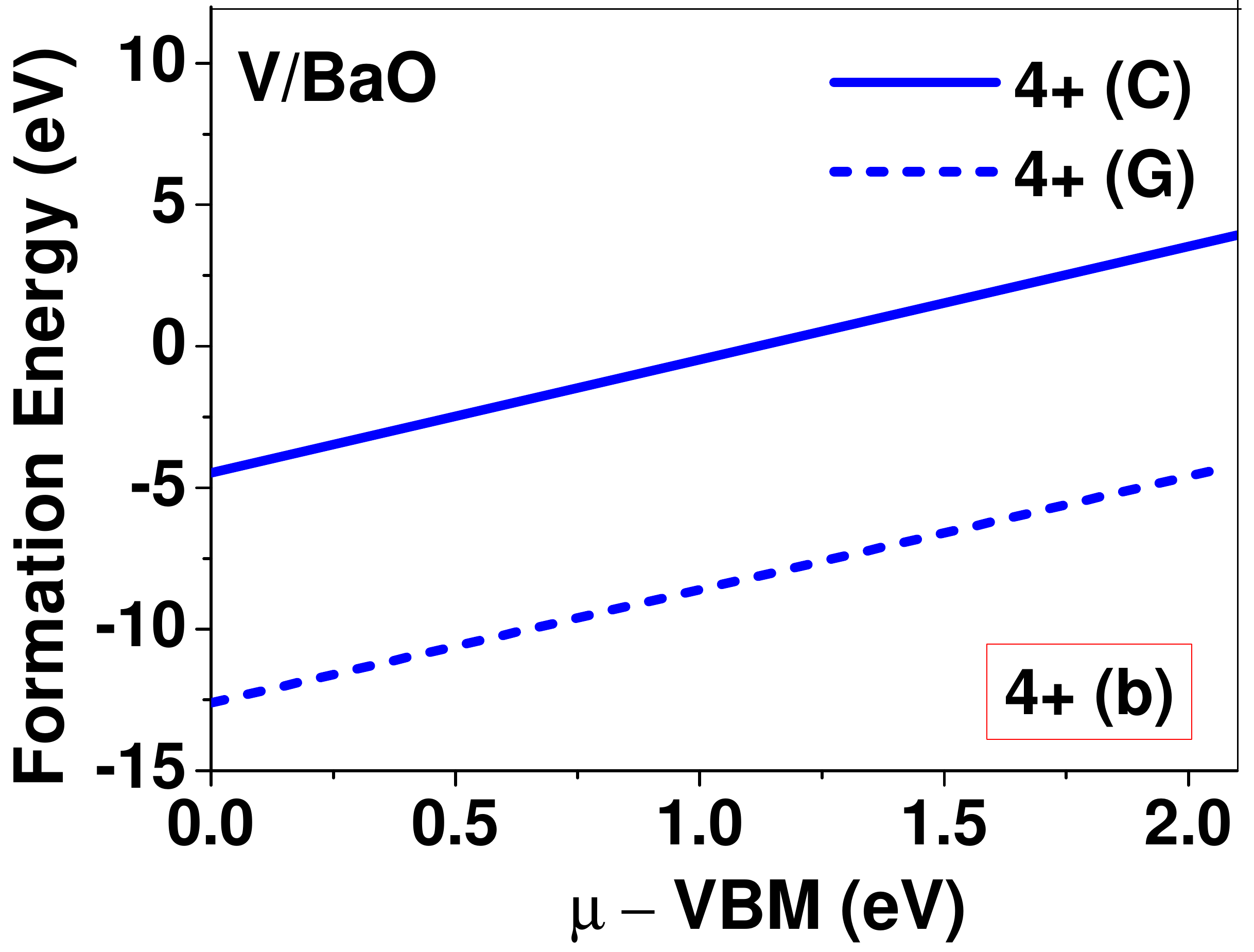}
    \end{subfigure}
    \hfill
    \begin{subfigure}
        \centering
        \includegraphics[height=1.31in]{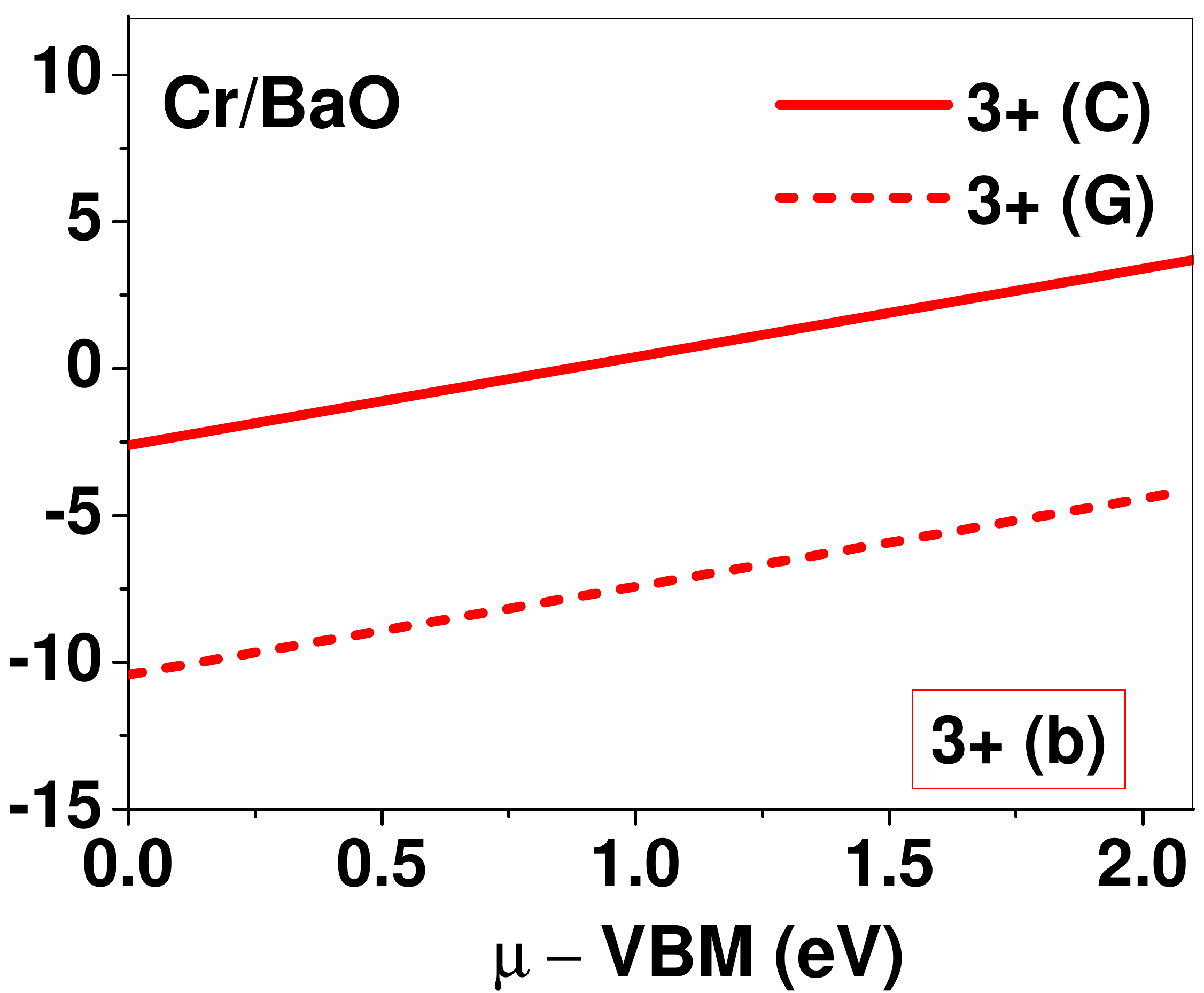}
    \end{subfigure}
    \hfill
    \begin{subfigure}
        \centering
        \includegraphics[height=1.3in]{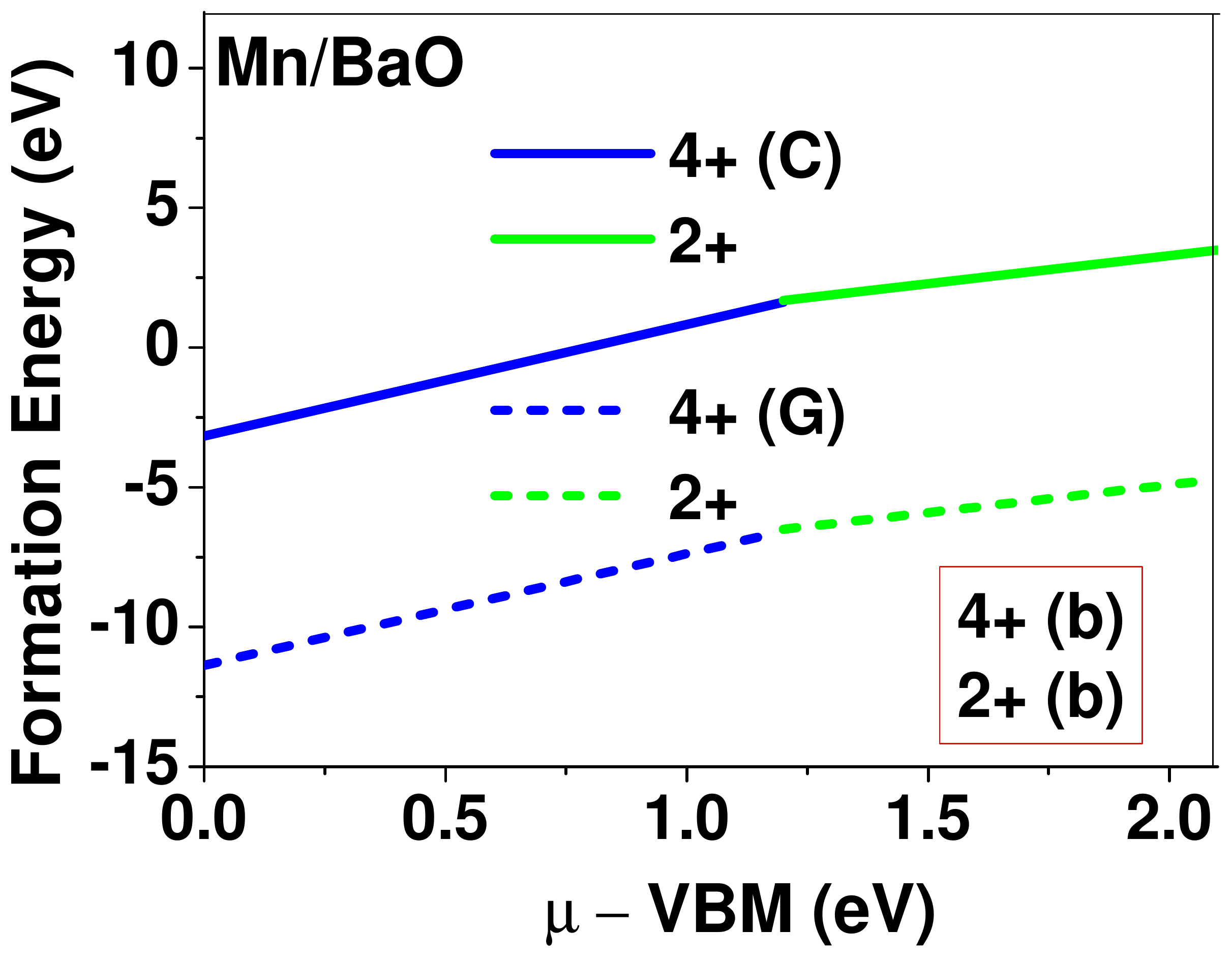}
    \end{subfigure}
    \hfill
    \begin{subfigure}
        \centering
        \includegraphics[height=1.31in]{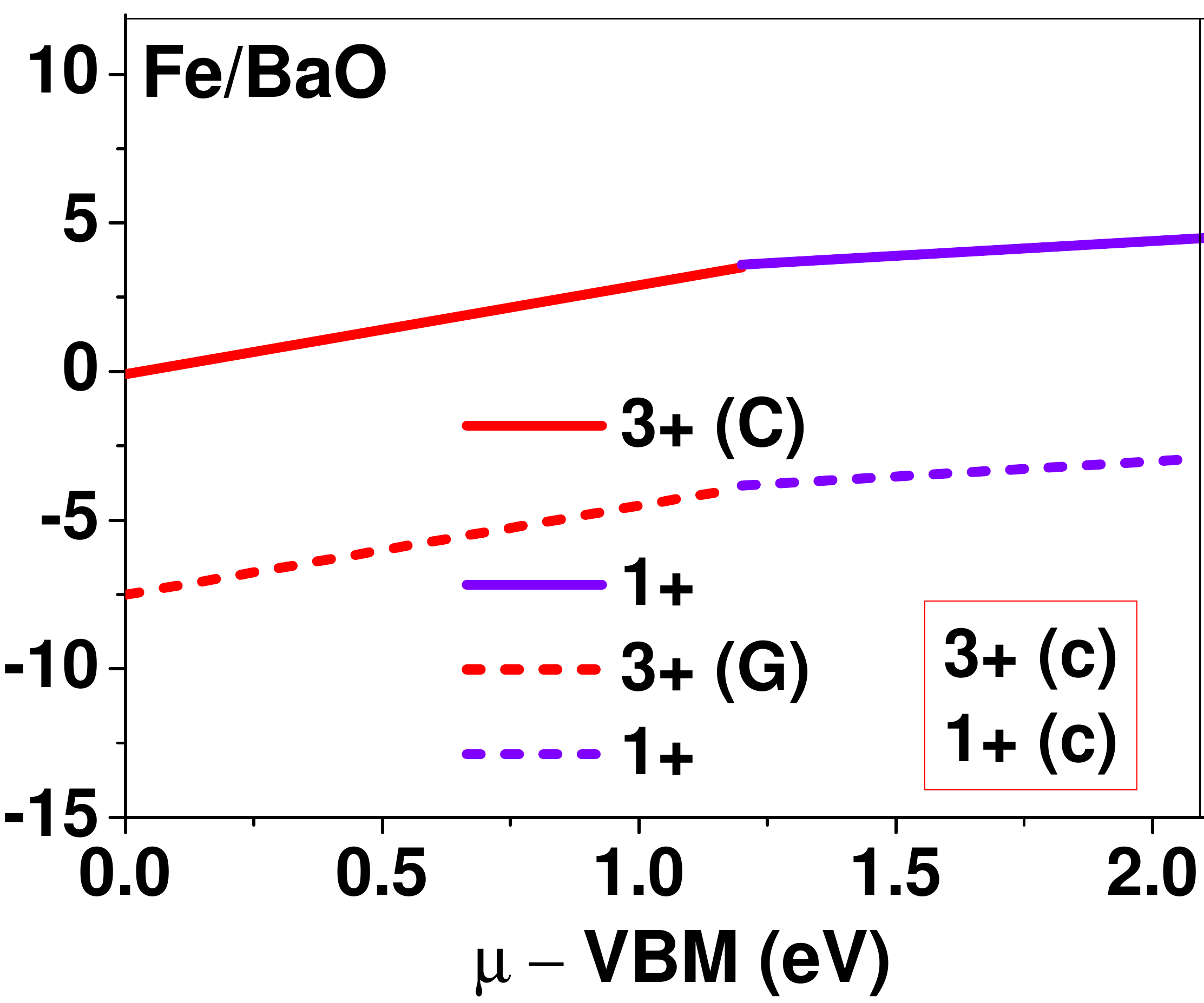}
    \end{subfigure}
    \hfill
    \begin{subfigure}
        \centering
        \includegraphics[height=1.3in]{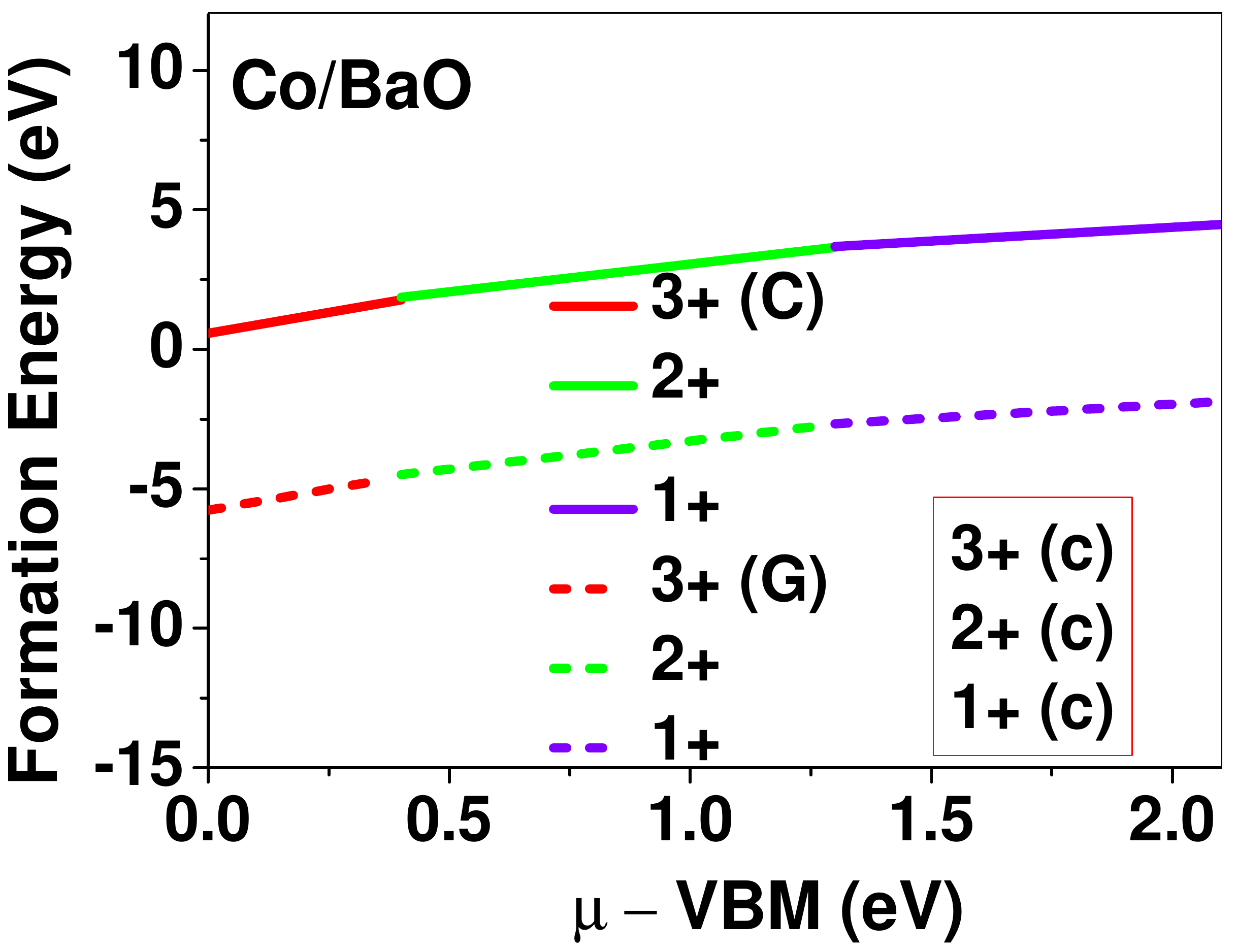}
    \end{subfigure}
    \hfill
    \begin{subfigure}
        \centering
        \includegraphics[height=1.31in]{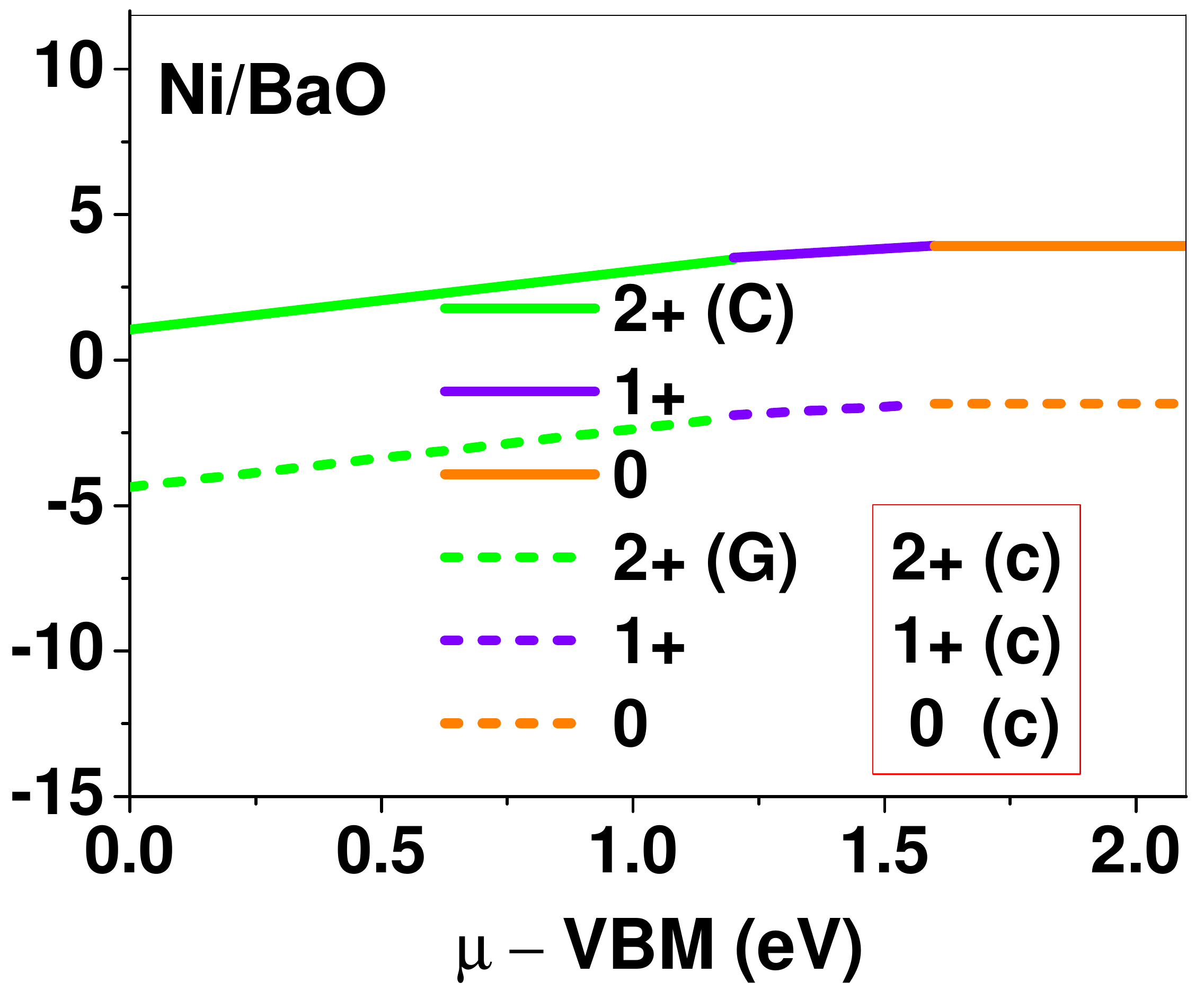}
    \end{subfigure}
    \hfill
    \begin{subfigure}
        \centering
        \includegraphics[height=1.3in]{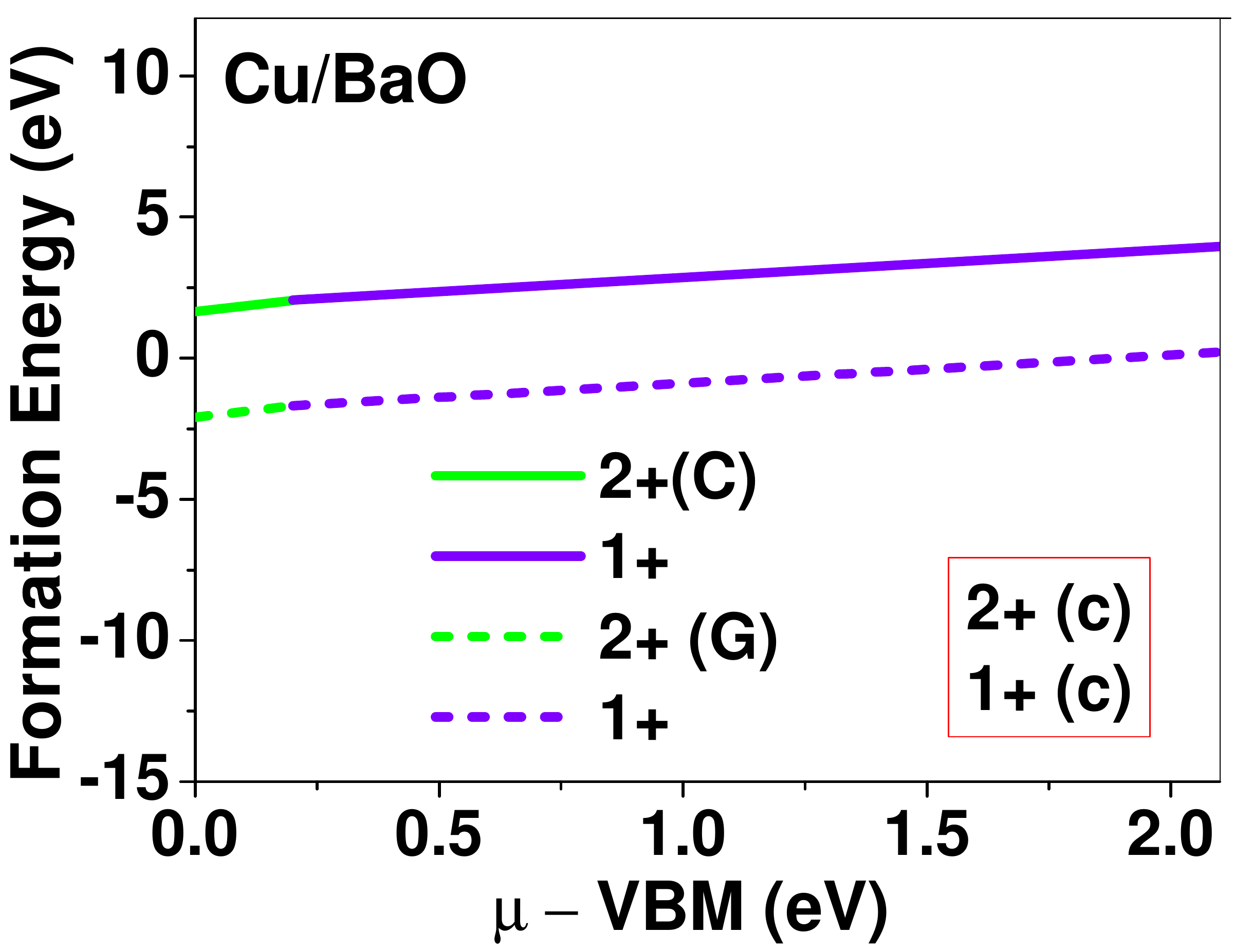}
    \end{subfigure}
    \hfill
    \begin{subfigure}
        \centering
        \includegraphics[height=1.31in]{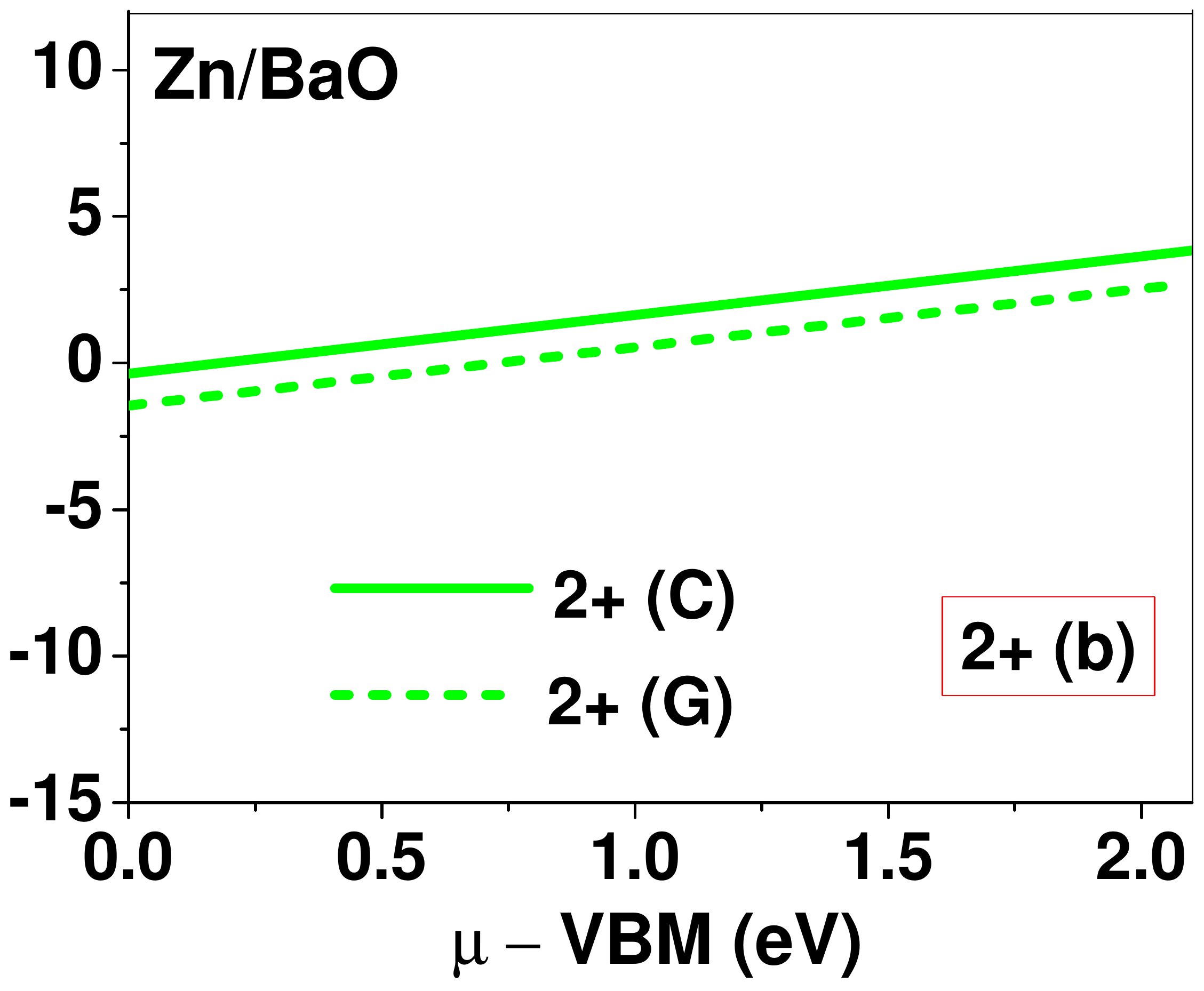}
    \end{subfigure}
    \caption{Formation energy for the neutral and charged dopants in BaO as a function of electronic chemical potential $\mu$; here $\mu$ is referenced to the valence band maximum (VBM). Dashed and solid lines represent the TM dopant formation energies, with gaseous(G) and crystalline(C) metal energy as references, respectively. The preferred defect site for each charge state is indicated in the bracket. Here (a-d) corresponds to Fig.~\ref{structure} (a-d) that shows the atomic structure of TM dopants in rock-salt oxides.} 
{\label{bao_form}}
\end{figure}

\begin{figure}[h!]
\includegraphics[width=2.8in]{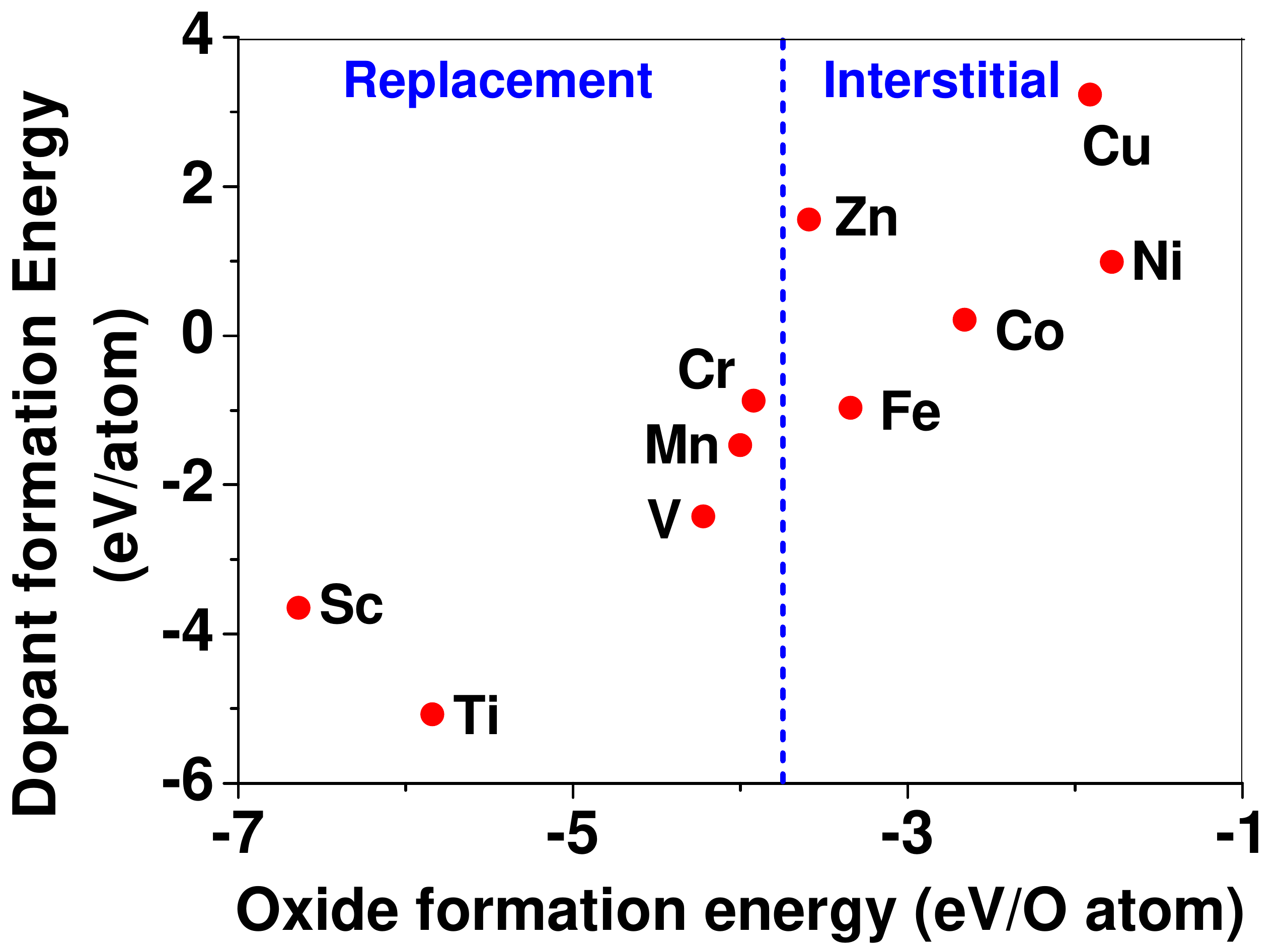}
\caption{Dopant formation energy of TM atoms (for crystalline energy reference) at $\mu$=VBM vs. experimentally measured enthalpy of formation of TM oxides per oxygen atom \cite{jain_prb,cox1989}). The vertical dashed line in the figure indicates the transition from replacement to interstitial as preferred defect site.}
{\label{enthalpy}}
\end{figure}

Dopant formation energy at preferred defect site as a function of electronic chemical potential $\mu$ for all the TM atoms in stable charge states is shown in Fig.~\ref{mgo_form} and Fig.~\ref{bao_form} for MgO and BaO, respectively. $\mu$ varies from VBM up to the band-gap of the host oxide, obtained from our DFT calculation. TM dopant formation energies, with both gaseous and crystalline metal energy as references are shown in the figures. Formation energy of dopants in MgO is relatively higher than BaO for all TMs at all electronic chemical potentials. Formation energy is lower for BaO compared to MgO at conduction band minimum (CBM) because DFT predicted band gap is smaller for BaO compared to MgO. A careful observation of the formation energy at VBM for the crystalline metal energy reference reveals that formation energy is lowest for Sc, followed by Ti, V, Cr, Mn, Fe, Zn, Co, Ni and that of Cu is highest. This trend can be understood if we consider the oxygen affinity of TMs. We define oxygen affinity of TMs as enthalpy of formation per oxygen atom of TMs oxides, as shown in Fig.~\ref{enthalpy}. Here we have considered the TM oxide showing the highest enthalpy of oxide formation per oxygen atom. Higher the oxygen affinity of TM, more the stability of the dopant. 

Elements like Sc, Zn that are known to show only one valency, assume only one charge state as dopant for the whole range of electronic chemical potential studied. While other elements that take multiple valencies such as Ti, V, Cr, Mn, Fe, Co, Ni and Cu, do not necessarily prefer all the charge states. For example Ti, V and Cr in BaO take only one charge state. Preferred charge state for a given electronic chemical potential would depend on various factors such as pressure exerted by dopant on the supercell and the position of the Fermi level. We will discuss these factors in greater detail in the subsequent sections. 

%Table for oxygen affinity

\subsection{Charge on dopants}
Excess charge in the charged supercell should be confined to TM atom alone as neither Mg nor Ba takes multiple valency. We have calculated atomic charges on Mg, Ba and TM atoms in defected and defect free supercells using Bader decomposition scheme on the total charge density for each case \cite{HENKELMANbader, bader2007, bader2009}, and the results are listed below in Table\ref{tab:bader}. Bader charge of Mg atom in a defect free MgO supercell is 1.66e and of Ba atom in a defect free BaO supercell is 1.37e. From the table it is evident that charge states of Mg and Ba atoms remain almost unaltered irrespective of the TM atom and their preferred defect site (interstitial or replacement) in the oxide. Maximum change in the charge state of Mg due to incorporation of dopant is 0.09e (in case of Sc$^{3+}$ and Ti$^{4+}$). However, for a given TM in a particular charge state, change in the charge state of Ba is more than that of Mg, which is in agreement with the fact that Mg shows higher affinity for oxygen than Ba.
%Table for charge transfer
\begin{table}
  \begin{center}
    \caption{Bader charge ($\it{q}$) of TM atoms in most stable charge states and Mg and Ba atoms in defect supercell. Suffixes $'$R$'$ and $'$I$'$ refer to TM atoms replacing host cations and occupying interstitial sites respectively. When TM is in interstitial site, we report average charge on four Mg or Ba atoms surrounding TM but when TM replaces Mg atom in MgO, charge on the resulting Mg interstitial is reported.}
    \label{tab:bader}
     \begin{tabular}{c|c|c|c|c|c|c|c}
      \multicolumn{4}{c|}{MgO} & \multicolumn{4}{c}{BaO} \\
     
       TM & \textbf{q(e)} & TM & \textbf{q(e)} & TM & \textbf{q(e)} & TM & \textbf{q(e)} \\ 
       atom & & atom & & atom & & atom  \\            
      
       \hline
        Sc$^{3+}_R$ & 1.87 & Ti$^{4+}_R$ & 2.03 & Sc$^{3+}_I$ & 1.76 & Ti$^{4+}_I$ & 1.87\\
      	Mg & 1.57  & Mg & 1.57 & Ba & 1.43 & Ba & 1.44\\
      	\hline 
      	V$^{4+}_R$ & 1.86  & Cr$^{3+}_R$ & 1.59 & V$^{4+}_I$ & 0.96 & Cr$^{3+}_I$ & 1.62\\
      	Mg & 1.58  & Mg & 1.58 & Ba & 1.30 & Ba & 1.20\\
      	\hline
      	Mn$^{4+}_R$ & 1.66  & Fe$^{3+}_I$ & 1.12 & Mn$^{4+}_I$ & 1.73 & Fe$^{3+}_I$ & 1.10\\
      	Mg & 1.58  & Mg & 1.66 & Ba & 1.51 & Ba & 1.47\\
      	\hline
      	Co$^{3+}_I$& 0.89 & Ni$^{3+}_I$ & 0.93 & Co$^{3+}_I$& 0.90 & Ni$^{2+}_I$ & 0.73\\
      	Mg & 1.69 & Mg & 1.66 & Ba & 1.47 & Ba & 1.44\\
      	\hline
      	Cu$^{3+}_I$& 0.64 & Zn$^{2+}_I$ & 0.82 & Cu$^{3+}_I$& 0.69 & Zn$^{2+}_I$ & 0.99\\
      	Mg & 1.69 & Mg & 1.66 & Ba & 1.46 & Ba & 1.46\\
  \end{tabular}
  \end{center}
\end{table}

\subsection{Role of stress induced by dopant on its stability}

\begin{figure}[h!]
\includegraphics[width=2.5in]{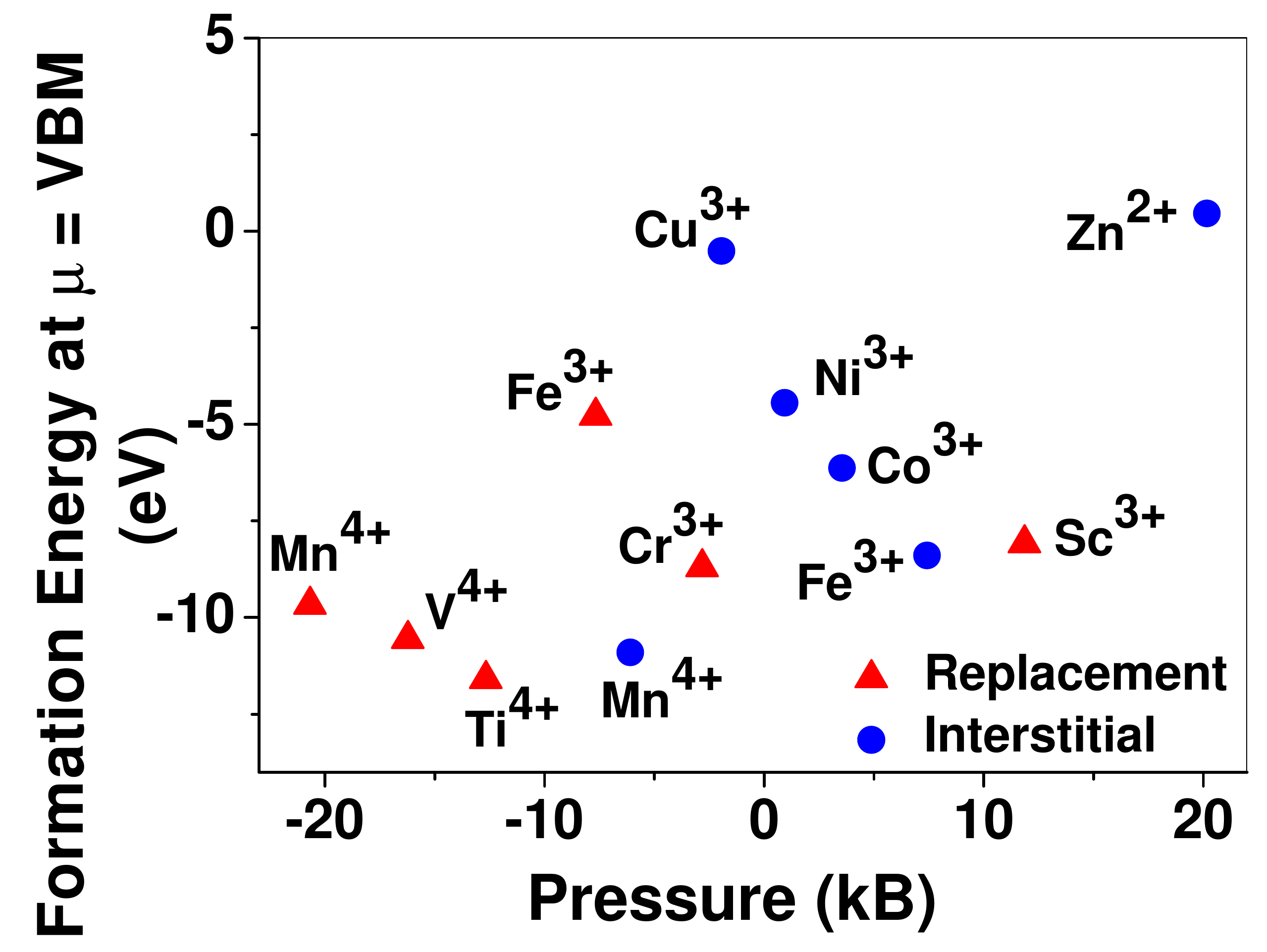}
\caption{Formation energy (with gaseous TM atom energy reference and at $\mu$=VBM) and corresponding hydrostatic pressure on the MgO supercell due to various TM dopants.}
{\label{mgo-pressure}}
\end{figure}

\begin{figure}[h!]
\includegraphics[width=2.5in]{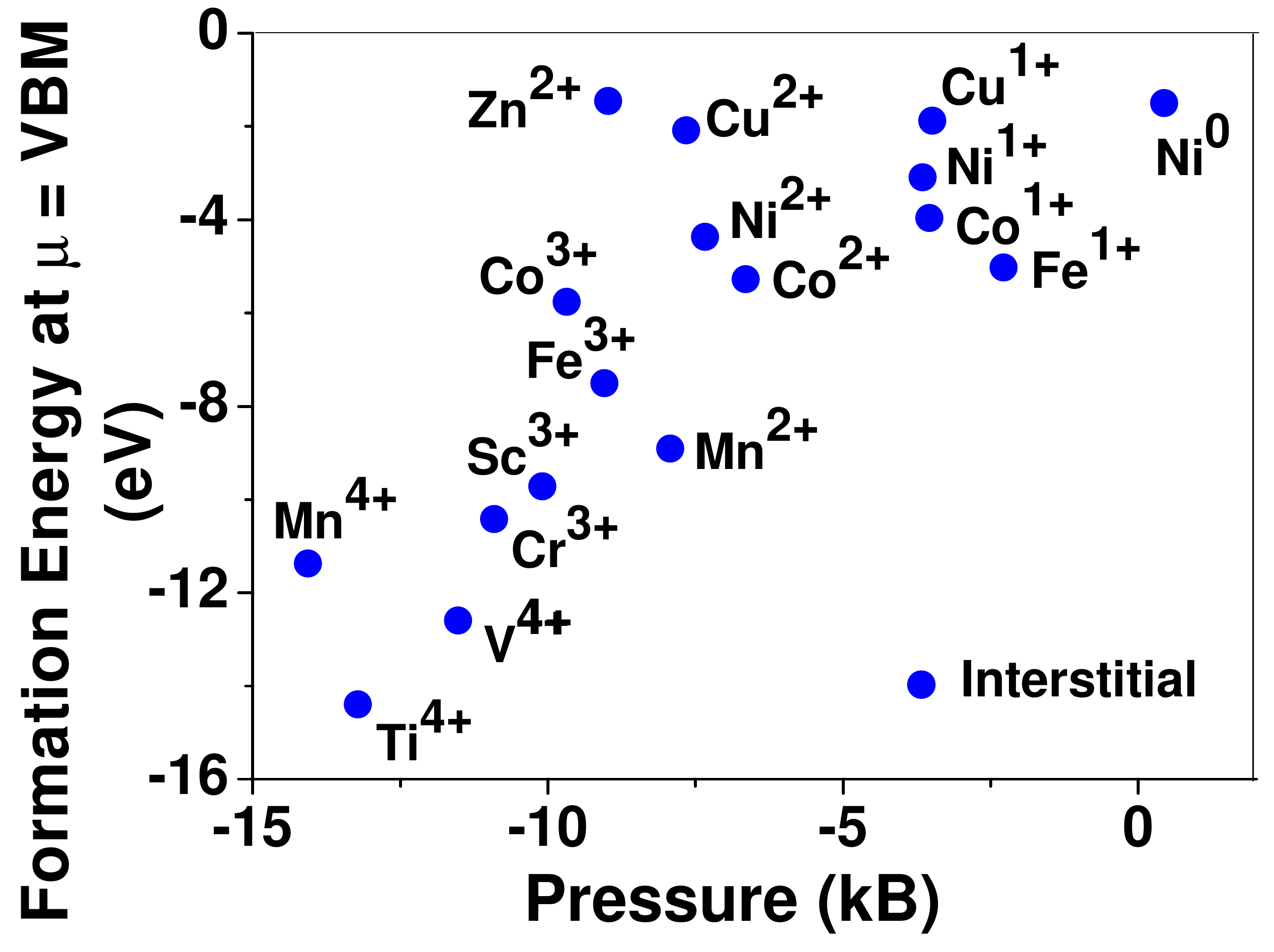}
\caption{Formation energy (with gaseous TM atom energy reference and at $\mu$=VBM) and corresponding hydrostatic pressure on the BaO supercell due to various TM dopants. All the TM dopants occupy interstitial sites in BaO.}
{\label{bao-pressure}}
\end{figure}

To understand the role of stress induced by dopant (on the host oxide) on its stability, we have calculated hydrostatic pressure exerted by the dopant on the supercell. Fig.~\ref{mgo-pressure} shows the hydrostatic pressure exerted by TM dopants on MgO supercell in their most stable charge states at $\mu$=VBM, and the corresponding defect formation energies. Fig.~\ref{bao-pressure} shows the hydrostatic pressure exerted by TM dopants in all stable charge states in BaO, and their dopant formation energies at $\mu$=VBM. If size of the dopant occupying interstitial site is larger than the available space then it exerts a positive pressure and if it is smaller then it exerts a negative pressure. For cases where TM atom replaces lattice Mg atom, pressure developed on the supercell is due to changes in both the oxygen octahedral volume surrounding TM ion and the oxygen tetrahedra surrounding the replaced Mg ion (now at interstitial site). In case of BaO, all the dopants sit at interstitials, and the pressure exerted by defects are mostly negative in nature. This observation is expected as interstitial space available in BaO is relatively larger than the size of dopants. As dopants get more positively charged, their size decreases, and hence pressure exerted on the supercell becomes more and more negative.

Although pressure exerted by dopant on the supercell is expected to play a substantial role in deciding stability of dopant but our results suggest that pressure alone does not explain formation energy and hence the site preference. Dopant formation energy of Cu$^{3+}$ is relatively higher than Cr$^{3+}$ and Fe$^{3+}$, while all of them exert very small pressure on the supercell. Similarly, Mn$^{4+}$ in interstitial site exerts small pressure compared to Mn$^{4+}$ replacing host cation, still replacement of lattice Mg atom by Mn is preferred. 

% MgO Density of States
\begin{figure}[t!]
    \centering
    \begin{subfigure}
        \centering
        \includegraphics[height=1.2in]{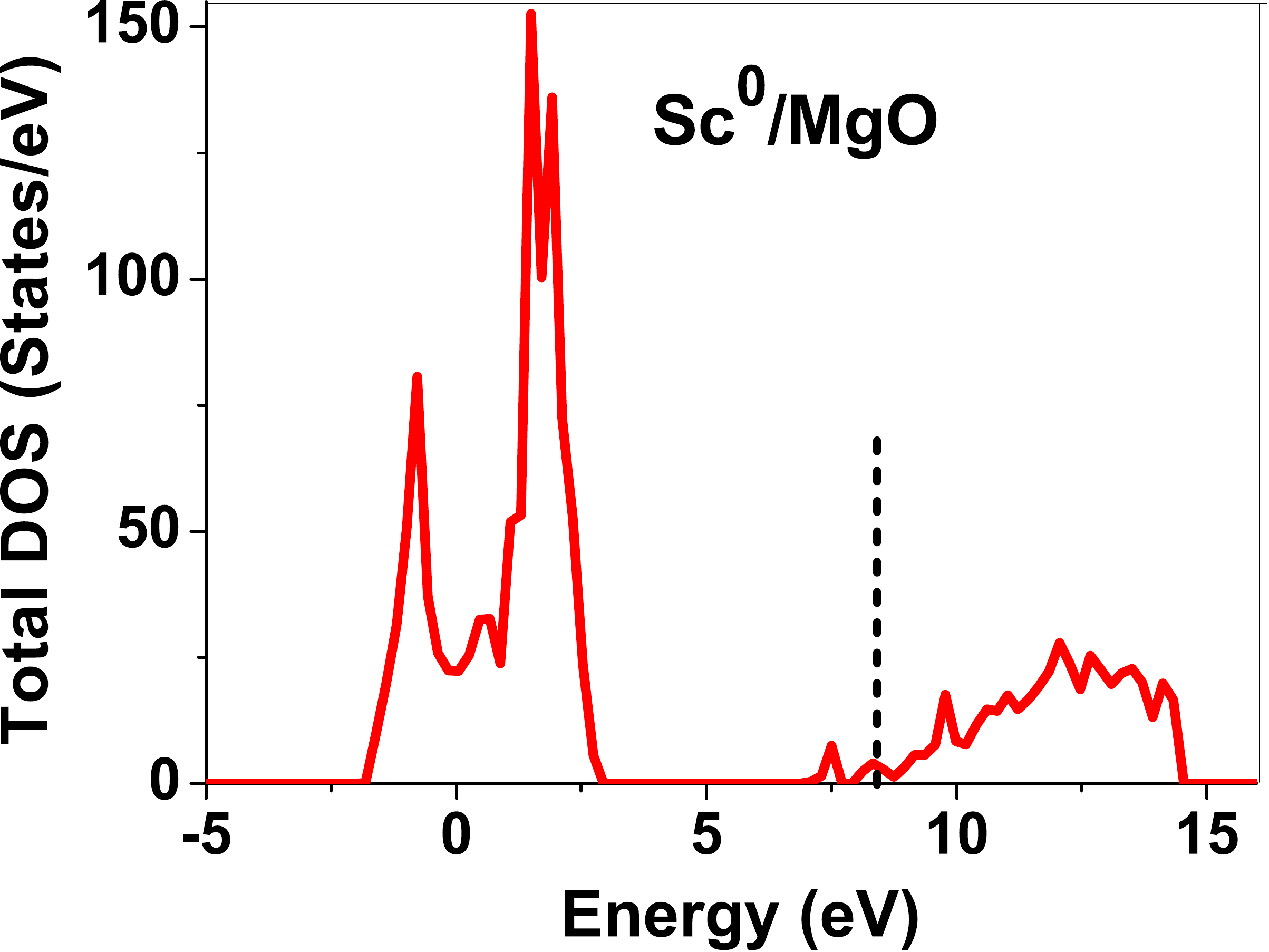}
    \end{subfigure}%
    \hfill
    \begin{subfigure}
        \centering
        \includegraphics[height=1.2in]{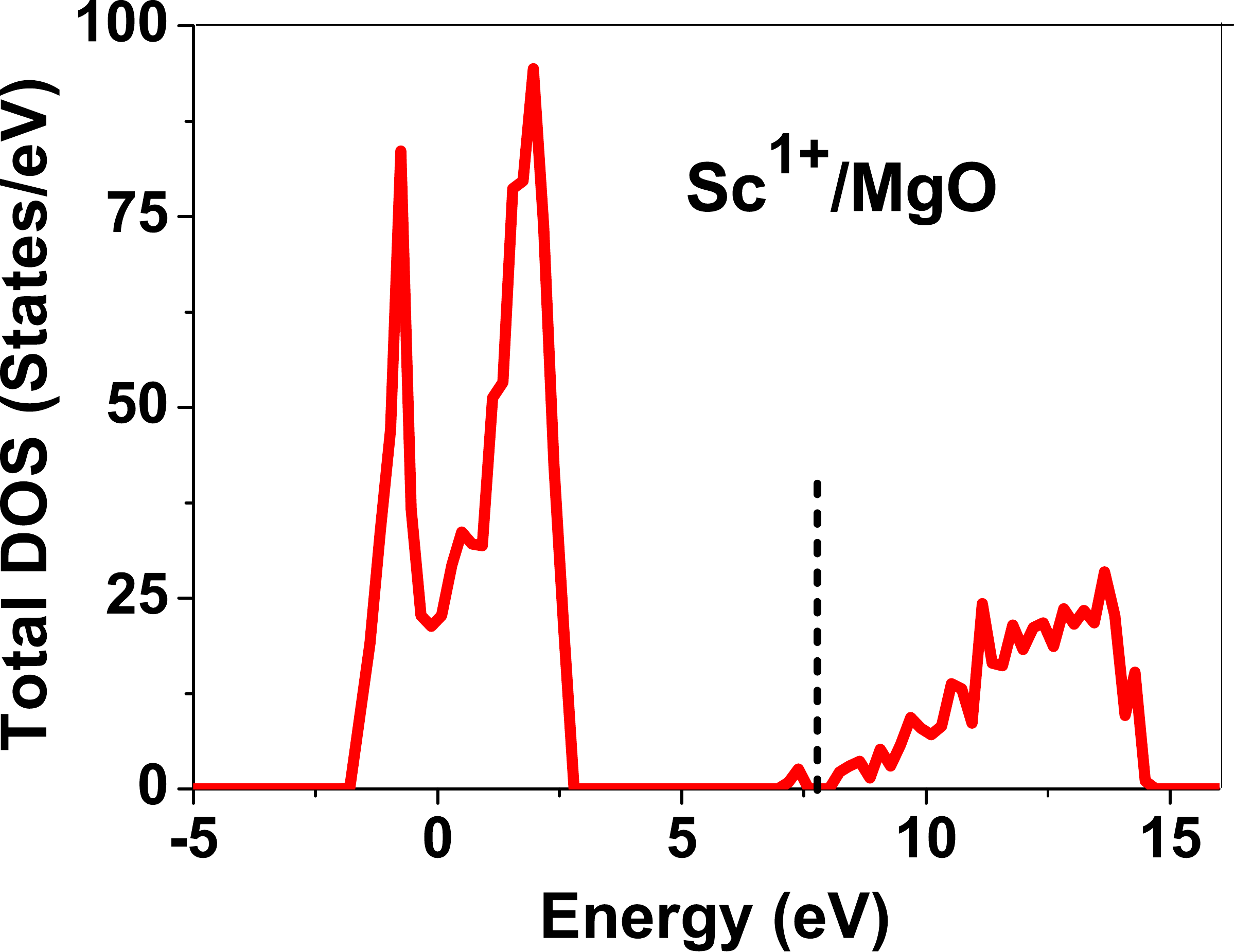}
    \end{subfigure}
    \hfill
    \begin{subfigure}
        \centering
        \includegraphics[height=1.2in]{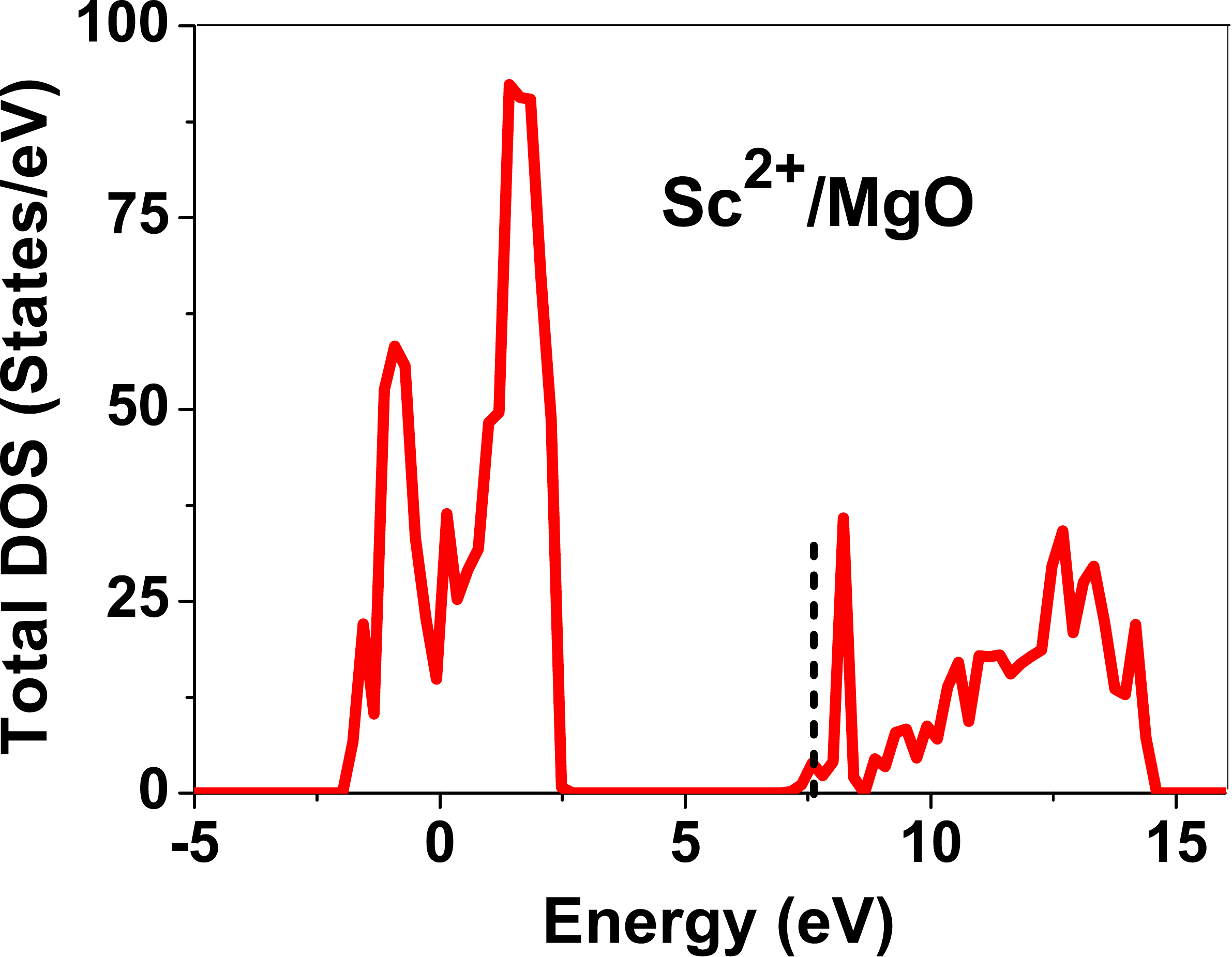}
    \end{subfigure}
    \hfill
    \begin{subfigure}
        \centering
        \includegraphics[height=1.2in]{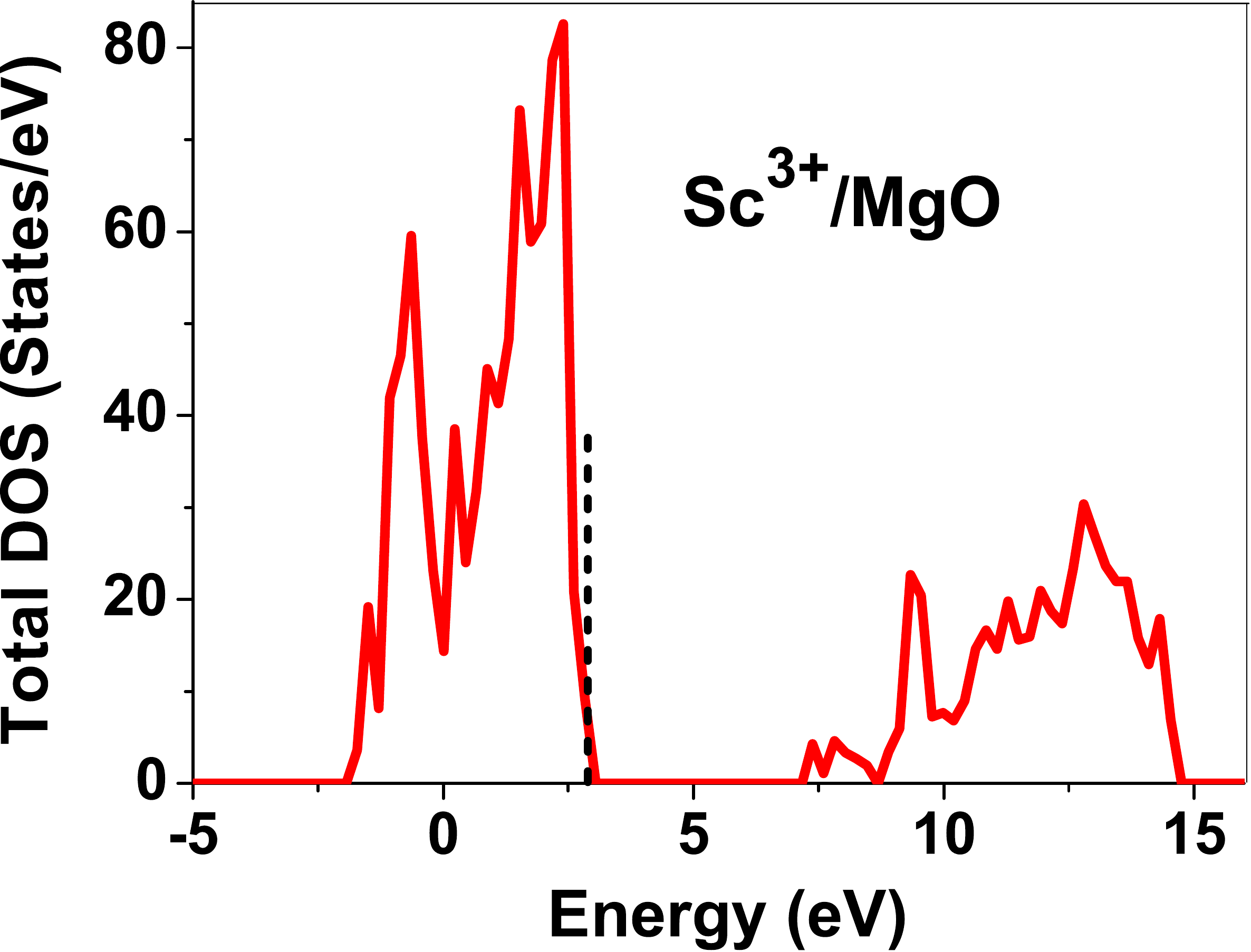}
    \end{subfigure}
    \hfill
    \begin{subfigure}
        \centering
        \includegraphics[height=1.2in]{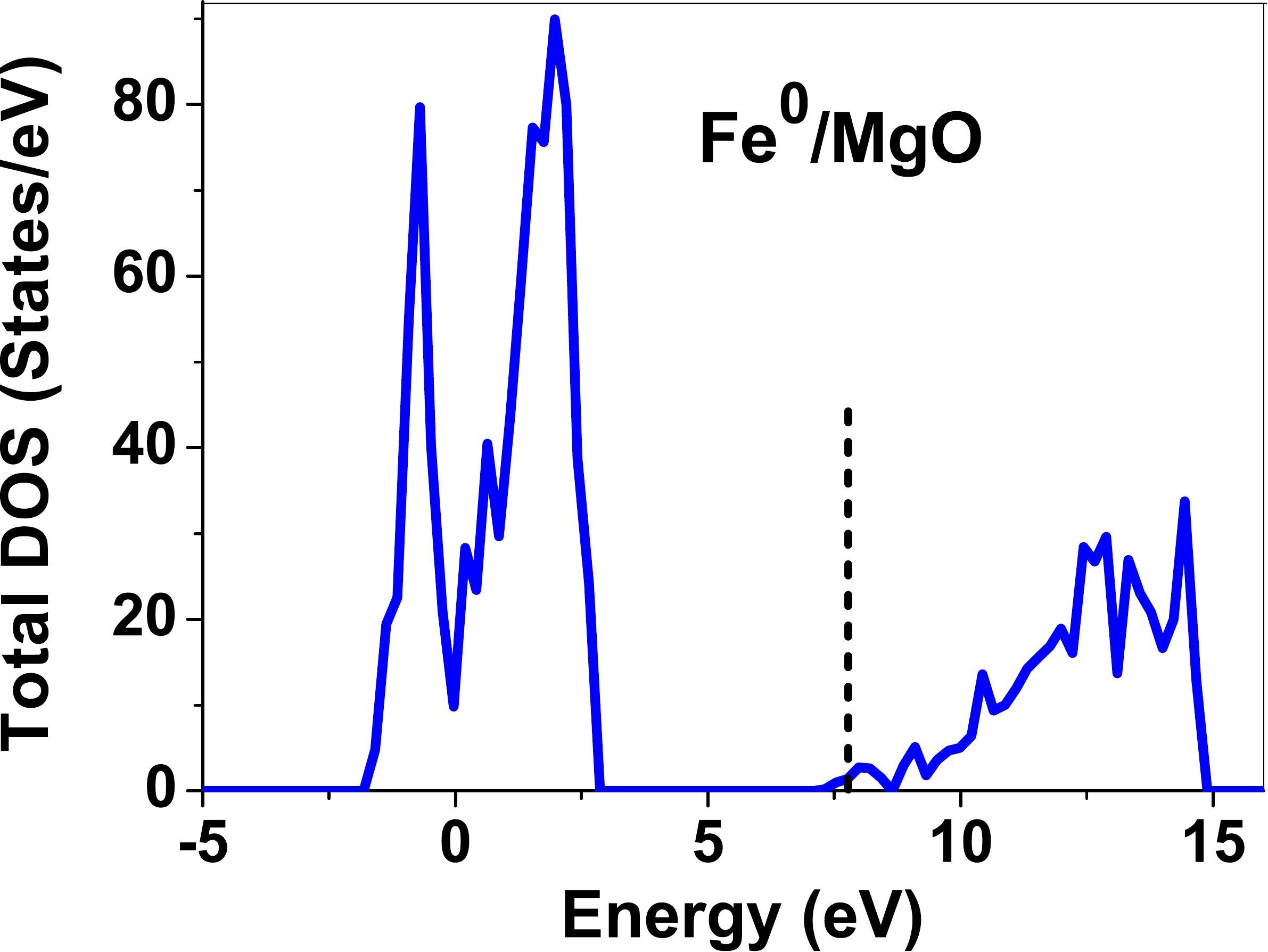}
    \end{subfigure}
    \hfill
    \begin{subfigure}
        \centering
        \includegraphics[height=1.2in]{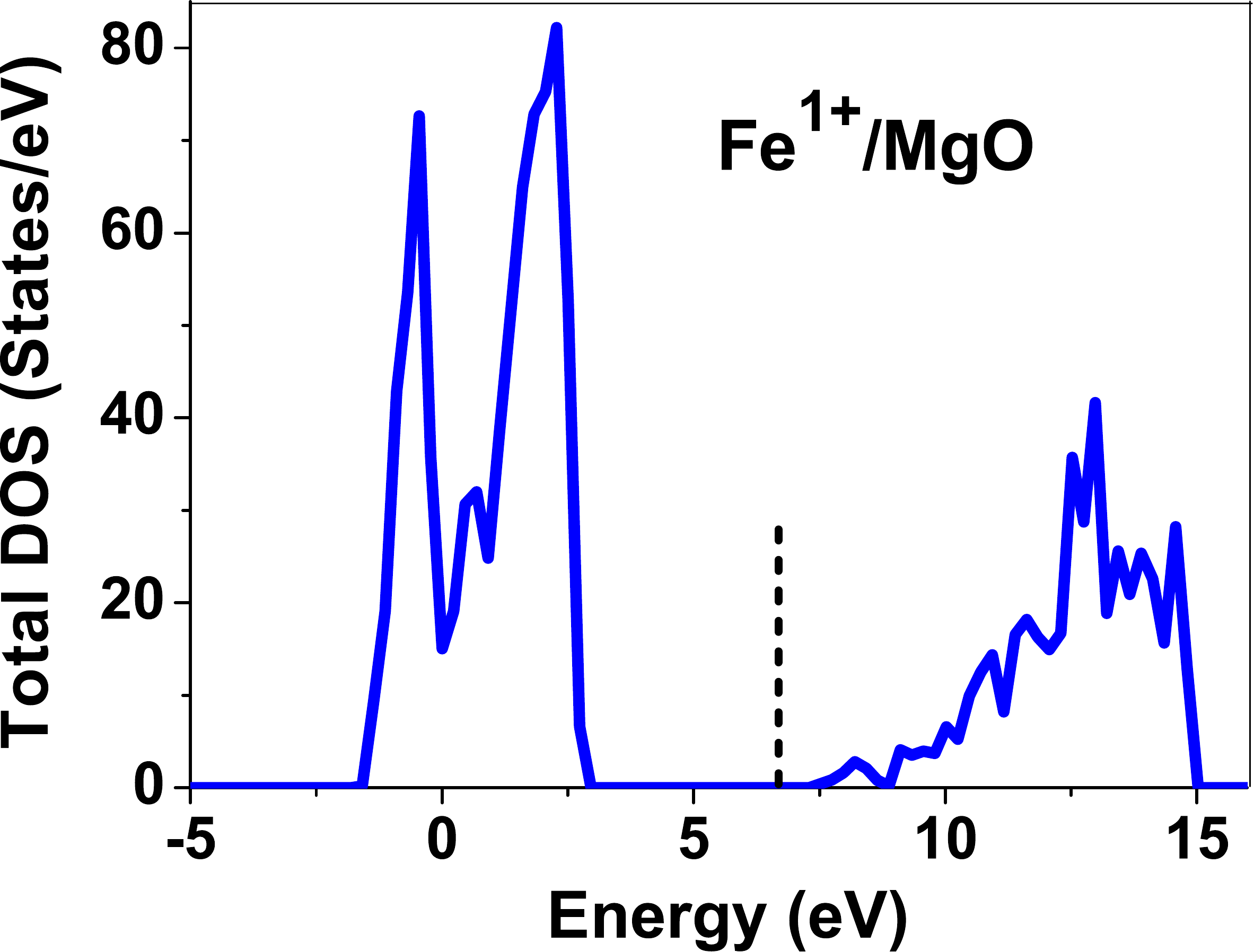}
    \end{subfigure}
    \hfill
    \hfill
    \begin{subfigure}
        \centering
        \includegraphics[height=1.2in]{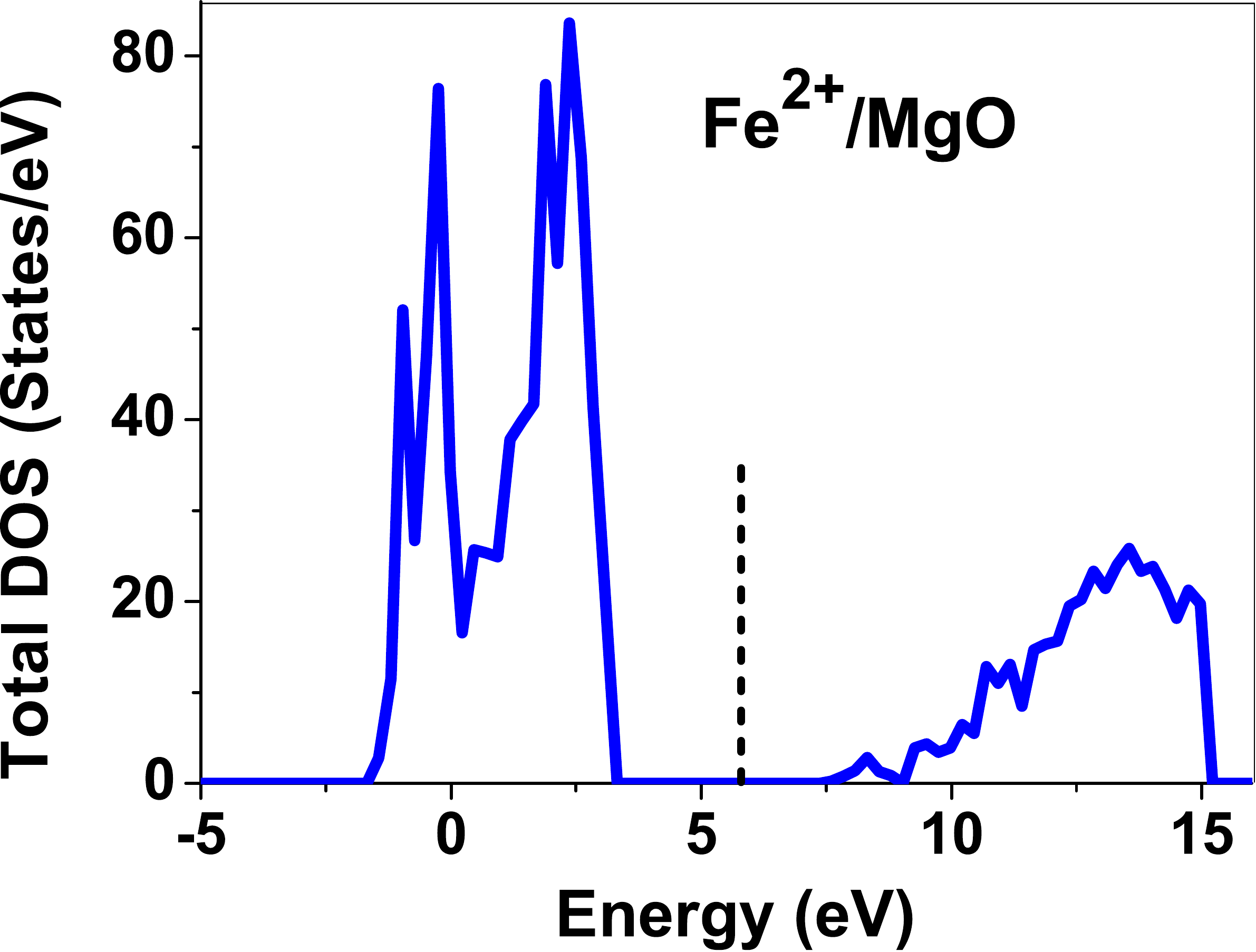}
    \end{subfigure}
    \hfill
    \begin{subfigure}
        \centering
        \includegraphics[height=1.2in]{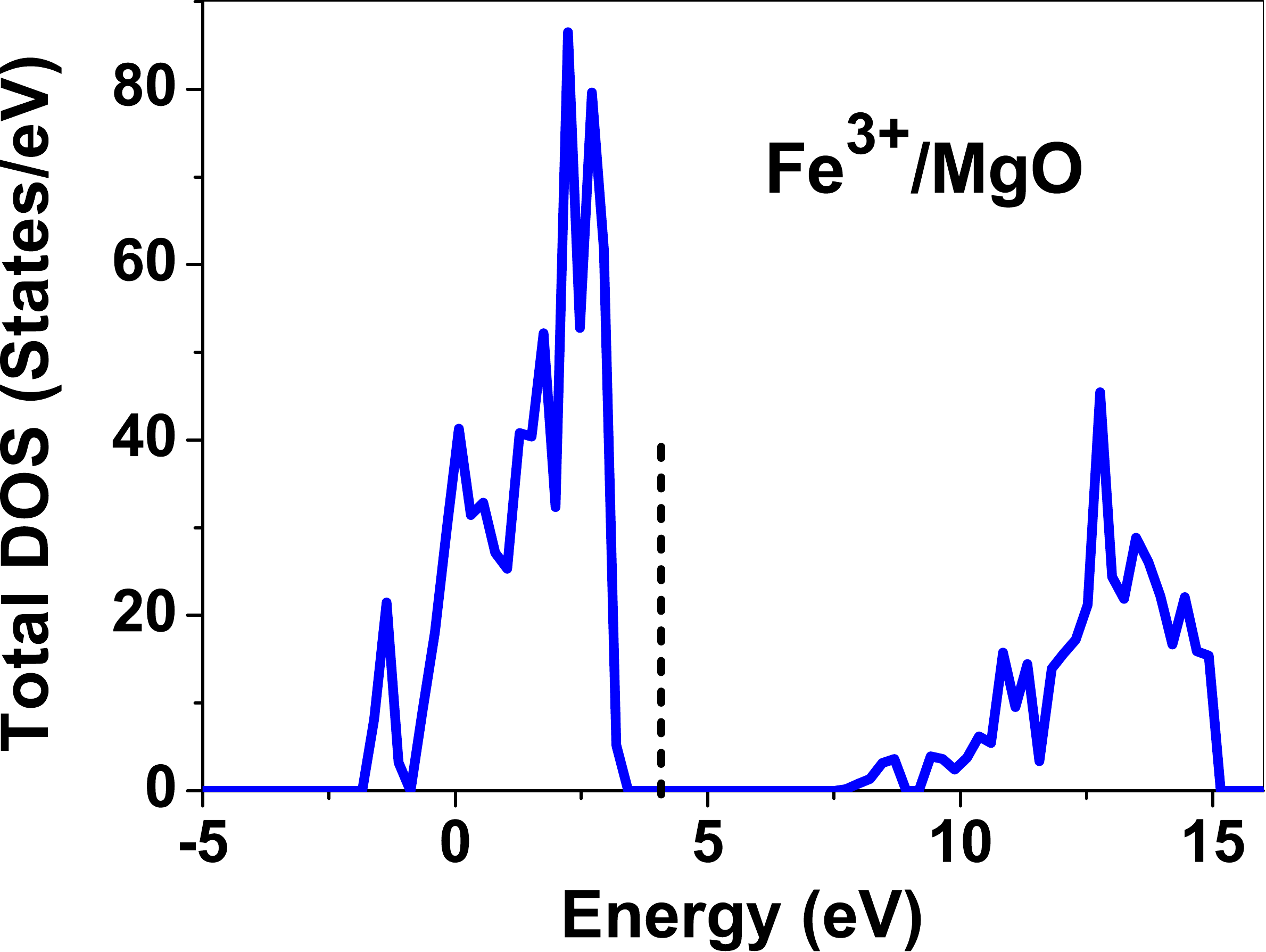}
    \end{subfigure}
    \hfill
    \begin{subfigure}
        \centering
        \includegraphics[height=1.2in]{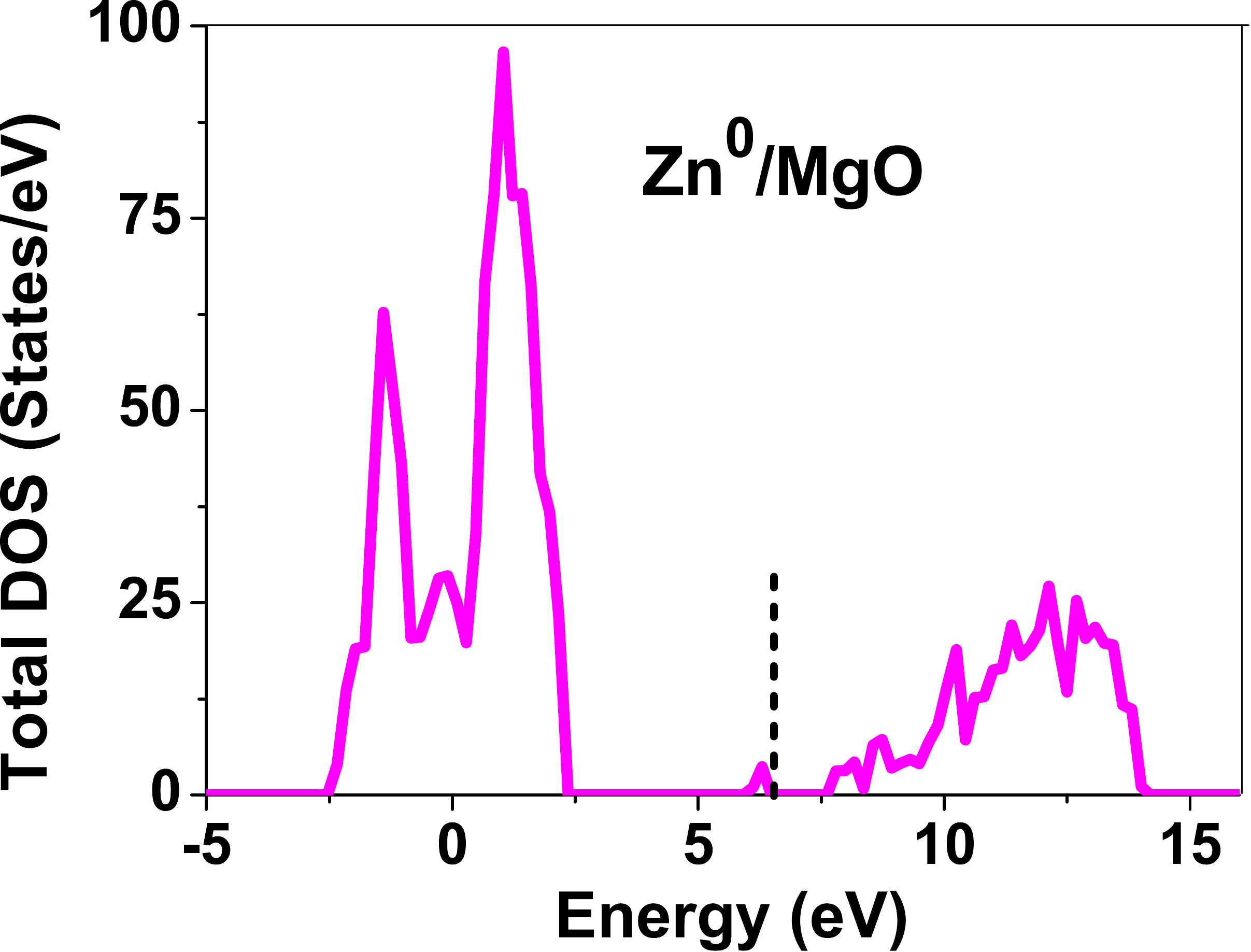}
    \end{subfigure}
    \hfill
    \begin{subfigure}
        \centering
        \includegraphics[height=1.2in]{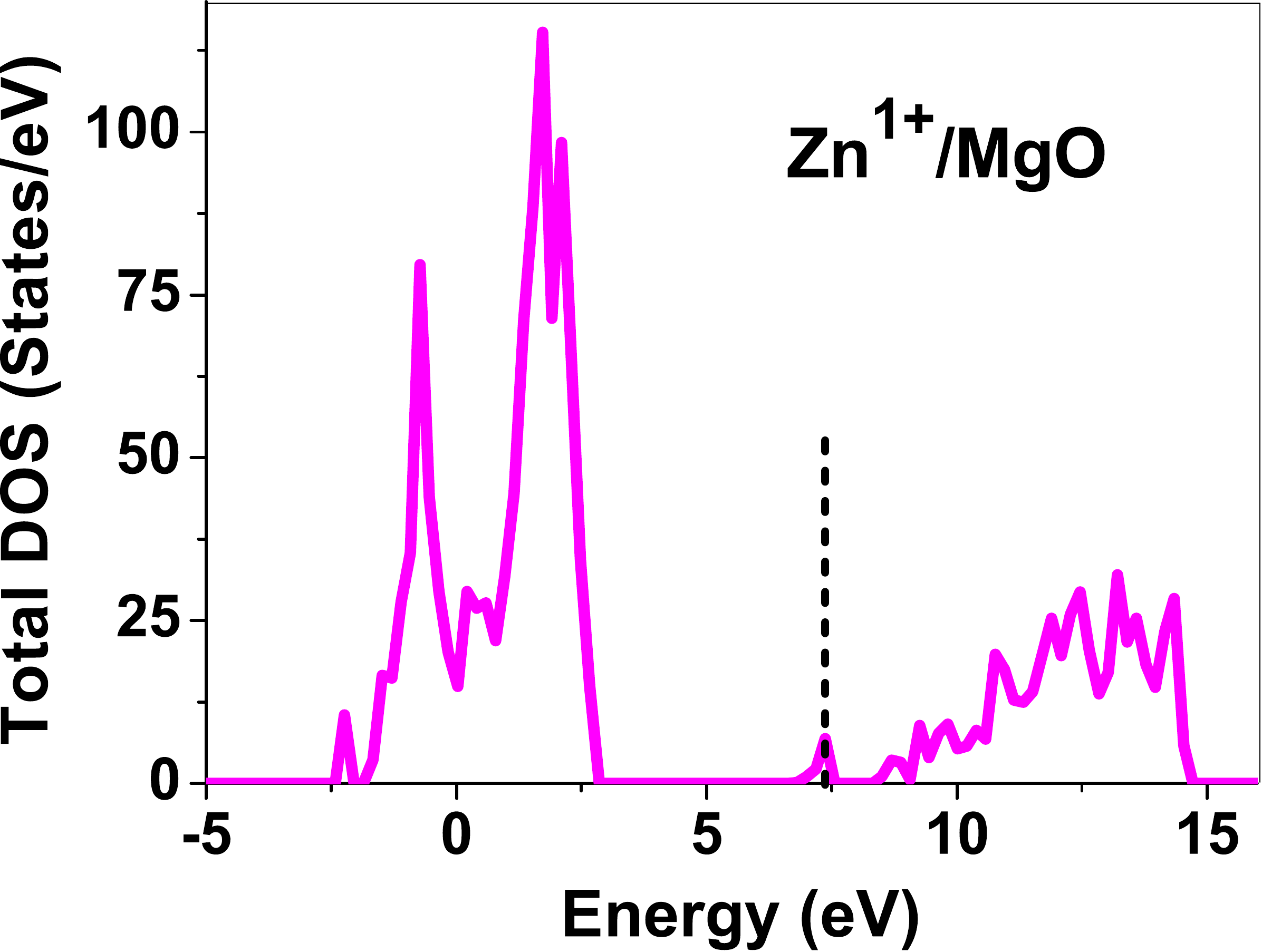}
    \end{subfigure}
    \hfill
    \begin{subfigure}
        \centering
        \includegraphics[height=1.2in]{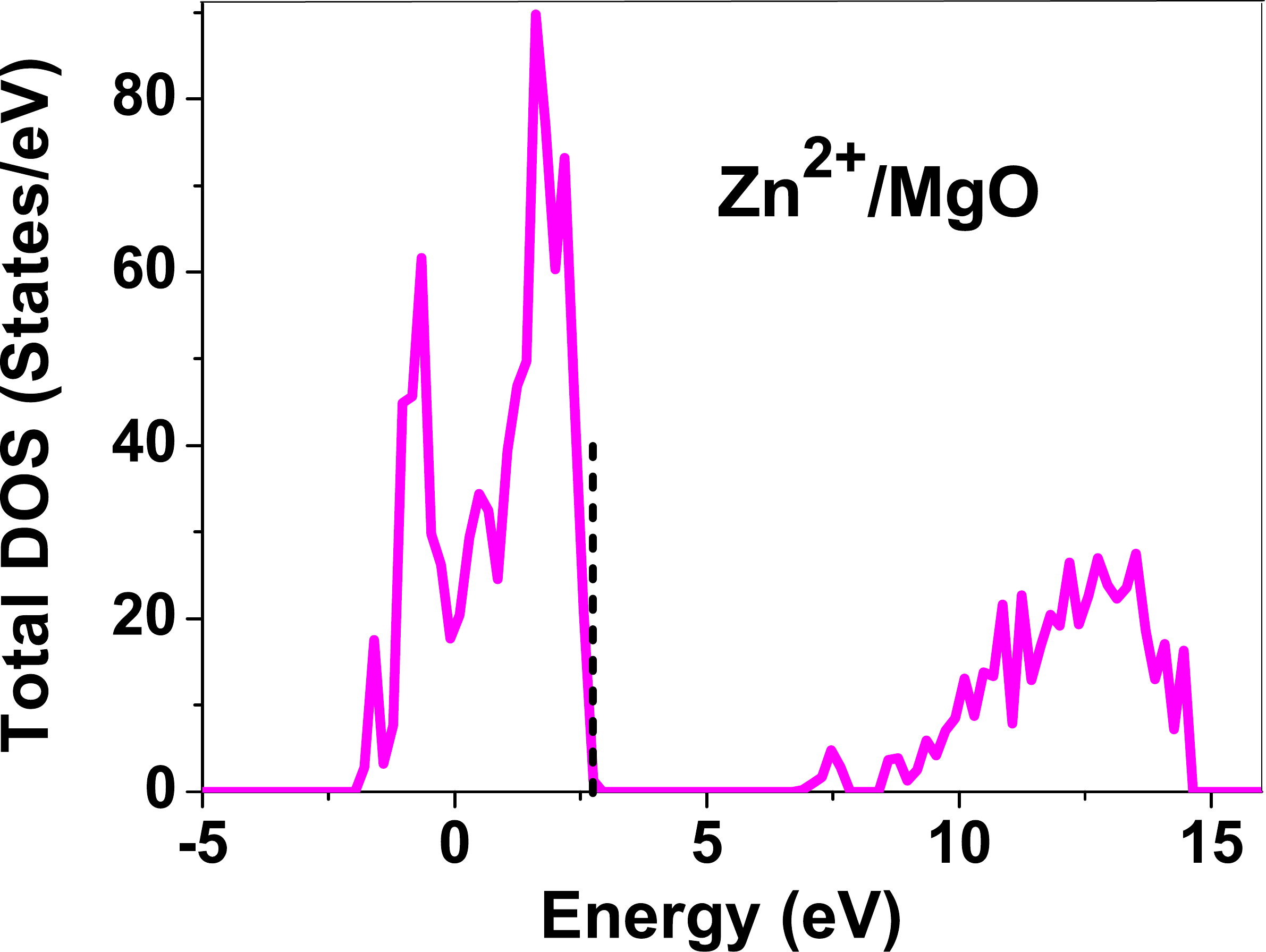}
    \end{subfigure}
    \hfill
    \begin{subfigure}
        \centering
        \includegraphics[height=1.2in]{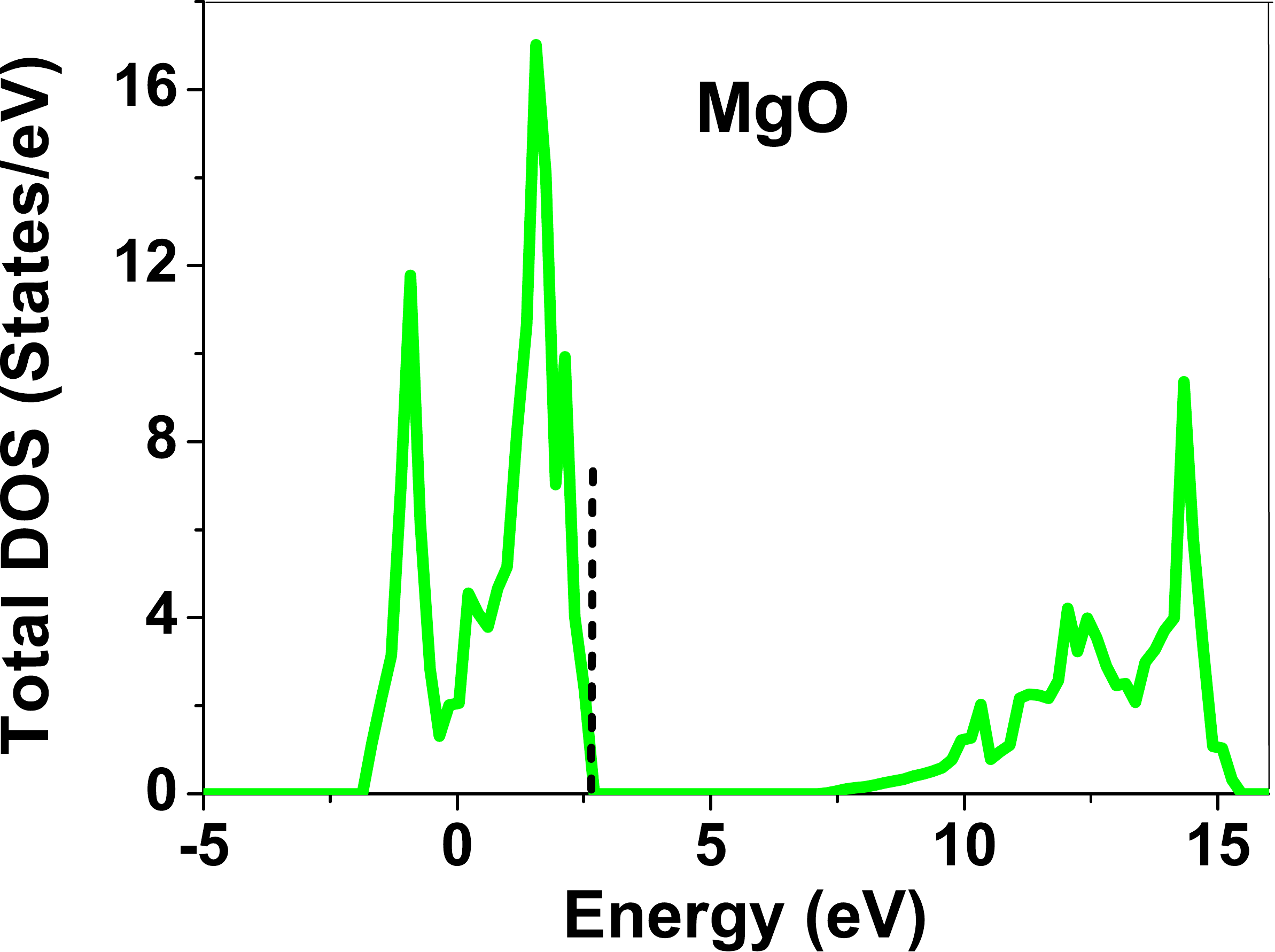}
    \end{subfigure}
    \caption{Total density of states of TM doped MgO and pure MgO. The vertical dashed line corresponds to the Fermi level. Electrostatic potentials of the defect supercell is aligned to that of pure MgO.} 
{\label{mgo_dos}}
\end{figure}

\subsection{Electronic structure}
As pressure exerted by a dopant is not sufficient to explain its stability, we consider the electronic structure of pure and doped MgO and BaO to get a better insight on the stability of dopants. We have calculated density of states (DOS) for the doped and undoped MgO and BaO supercells. In Fig.~\ref{mgo_dos} we plot DOS along with the Fermi level for few cases of doped MgO: Sc at replacement site, and Fe and Zn at interstitial site, with an aim to provide rationale for preferred charge states. These three cases are representative of TM dopant at: 1) replacement site with single valency (Sc), 2) interstitial site with multiple valency (Fe), and 3) interstitial site with single valency (Zn). Position of Fermi level with respect to VBM and CBM is a good indicator of relative stability of various charge states of a system. Closer the Fermi level is to VBM, more stable is the system ~\cite{jhxu_1987}.

For Sc$^0$, Sc$^{1+}$ and Sc$^{2+}$ in MgO, the Fermi level is in the anti-bonding region, while that of Sc$^{3+}$ is in the bonding region, which shows that Sc in 3+ charge state is the most stable state in MgO compared to its other charge states. A similar trend has been observed for Zn in MgO also; the Fermi level for Zn$^0$ and Zn$^{1+}$ lies in the anti-bonding region of the density of states while for Zn$^{2+}$, it is in the bonding region. Hence, Zn is most stable in 2+ charge state in MgO. It is interesting to note that for both the cases no defect state appears in the band gap. Both of them are known to take only one valency and they are found to be stable in that valency alone. In contrast, Fe can take multiple charge states and as dopant it is found to be stable in multiple charge states. This could be attributed to defect states appearing in the bandgap.
 
%Table for ionic radius
\begin{table} 
  \begin{center}
    \caption{Shannon radii for TM atoms, Mg and Ba in 6C environment~\cite{shannon}}
    \label{tab:table1}
    \begin{tabular}{c|c|c|c} 
     \textbf{Atom} & \textbf{Ionic radius} & \textbf{Atom} & \textbf{Ionic radius}\\
       & (\AA) &  & (\AA)\\ 
      \hline        
      \textbf{Mg$^{2+}$} & 0.72 & \textbf{Ba$^{2+}$} & 1.35\\    
      Sc$^{3+}$ & 0.75 & Fe$^{3+}$ & 0.65\\
      Ti$^{4+}$ & 0.61 & Fe$^{2+}$ & 0.78\\
      Ti$^{3+}$ & 0.67 & Co$^{3+}$ & 0.61\\
      Ti$^{2+}$ & 0.86 & Co$^{2+}$ & 0.75\\
      V$^{4+}$ & 0.58 & Ni$^{3+}$ & 0.60\\
      V$^{3+}$ & 0.64 & Ni$^{2+}$ & 0.69\\
      Cr$^{3+}$ & 0.61 & Cu$^{3+}$ & 0.54\\
      Mn$^{4+}$ & 0.53 & Cu$^{2+}$ & 0.73\\
      Mn$^{3+}$ & 0.65 & Cu$^{1+}$ & 0.77\\
      Mn$^{2+}$ & 0.83 & Zn$^{2+}$ & 0.74\\
      \end{tabular}
  \end{center}
\end{table}

\subsection{Predicting site preference}
Relative ionic radius of the host cation and the TM dopant could be important in predicting whether TM would substitute cation or remain in interstitial site. If ionic radius of TM is smaller than host cation, it should prefer interstitial site, while a TM atom with higher ionic radius should substitute the host cation. We rely on Shannon ionic radii~\cite{shannon} to assess importance of ionic radius in predicting the site preference. As cations (Mg and Ba) in host oxides are coordinated by 6 oxygen atoms, Shannon ionic radii with 6 coordination are considered and are listed in Table~\ref{tab:table1}. It should also be noted that ionic radius of a TM dopant depends on the charge state of the dopant. For example, ionic radius of Mn varies from 0.53 \AA to 0.83 \AA, as the charge state changes from +4 to +2. Ionic radius of Ba$^{2+}$ is 1.35 \AA, which is significantly larger than the ionic radius of Ti$^{2+}$ (0.86 \AA), the largest dopant among all the TM dopants considered in our computation. Hence, in case of BaO all the dopants prefer interstitial sites. However, ionic radius of Mg$^{2+}$ (0.72 \AA) is comparable to all the dopants considered, as listed in Table~\ref{tab:table1}. While Zn, Cu, Ni, Co, Fe prefer to be in the interstitial sites in MgO, Mn, Cr, Sc, and Ti prefer to substitute Mg atoms. Transition from interstitial to substitution occurs from Fe to Mn. Ionic radius of Mn is both higher and lower than the ionic radius of Mg$^{2+}$ depending on its charge state, but it still prefers substitutional site. On the other hand, both Fe$^{2+}$ and Fe$^{3+}$ prefer to occupy the interstitial sites although ionic radius of Fe$^{2+}$ (0.65 \AA) is smaller and Fe$^{3+}$ (0.78 \AA) is larger than the ionic radius of Mg$^{2+}$. Thus it seems that when ionic radius of TM dopant is comparable to that of the cation, it has a minimal role in dictating the site preference. Instead, oxygen affinity (defined as enthalpy of formation of oxide per oxygen atom) serves as a better indicator for site preference. Fig.~\ref{enthalpy} shows dopant formation energy as a function of enthalpy of oxide formation per oxygen atom of TMs. As oxygen affinity increases (more negative enthalpy of oxide formation) replacement is more favoured than interstitial site. 

Based on our results, we propose an empirical rule for site preference of dopants in rock-salt oxides. It states that site preference is governed by both ionic radius and oxygen affinity of TM; if ionic radius of host cation is large, then TM occupies interstitial site as in case of BaO. However, if ionic radius of host cation is similar to that of TM, as in case of MgO, then TM replaces the cation provided oxygen affinity of TM is similar to that of the host cation, otherwise it occupies the interstitial site. This is in contrast to Pauling's first rule that takes only ionic radius of oxygen into account. 

\subsection{Implications for experiments}
Our results explain experimental observation of site preference of implanted Fe and Ni ions in MgO. Experiment shows that Ni atoms implanted in MgO at room temperature get distributed in the matrix but upon annealing Ni precipitates out with an average particle size of 8-10 nm \cite{ZHU2006}. This could be possible if implanted Ni atom occupies interstitial site as suggested by our calculations. Although, Ni prefers interstitial site, formation energy of the dopant with crystalline energy reference (shown by the solid line in Fig.~\ref{mgo_form}) is always positive for the entire range of electronic chemical potential. This suggests that Ni could occupy interstitial site but it is unstable in MgO. In order to estimate if Ni could precipitate out at high temperature (900$^0$C), as observed experimentally, we have calculated the barrier of transition from one interstitial site to the other, using nudged elastic band (NEB) method. Barriers for transition of Ni in MgO is very low for all its charge states; the lowest being ~0.02 eV for neutral Ni dopant in MgO. Migration of interstitial Mg atoms in MgO has been reported to have a barrier of 0.32 eV and have been shown to diffuse at the time scale of nanosecond at 300 K \cite{Uberuagaprl2004}. Hence Ni at interstitial with a barrier height as low as 0.02 eV would diffuse even faster at higher (900$^0$C) temperature and will cause Ni precipitation. 

Ni presents an interesting case to compare our way of calculating preference for interstitial site or substitutional site with the one where Mg atom is replaced with Ni atom in the supercell and the replaced Mg atoms go to metallic Mg, \cite{prada2012} commonly referred as substitutional formation energy. We consider neutral Ni as a defect in a 32 formula unit of MgO. Substitutional formation energy is defined as
\begin{equation} 
E_f=E_D+\mu_{Mg}-E_B-\mu_{Ni}
\end{equation}
Here E$_D$ and E$_B$ are total energies of the supercell containing Ni defect, and defect free MgO supercell respectively. $\mu_{Ni}$ and $\mu_{Mg}$ refer to the chemical potential (crystalline energy reference) of Ni and Mg, respectively. The dopant formation energy using the above equation came out to be 4.42 eV which is nearly 3.81 eV lower than the formation energy of Ni as interstitial in MgO, which would lead one to conclude that Ni would substitute Mg. This example points towards the importance of calculating interstitial and replacement formation energies as defined in this work in order to explore interstitial as a possible defect site for the TM atoms. 

Existing experiments\cite{WHITE1989} show that Fe implanted in MgO is stable in Fe$^{3+}$, Fe$^{2+}$ and Fe$^0$ charge states, but do not comment on where Fe ions sit in the host lattice. Here we show that Fe atoms occupy interstitial sites in MgO. Our calculations revealed that stable charge states for Fe in interstitial site are +3, +2 and 0, which is in good agreement with the experimental observation of stable charge states of Fe in MgO \cite{WHITE1989}. Substitutional defect formation of Fe in MgO show +3, +2, +1 and 0 as stable charge states of Fe, as has also been reported earlier \cite{larico_2013}. As charge states of substitution doped Fe do not match with experimental observation, hence Fe atoms most likely occupy interstitial sites in MgO \cite{WHITE1989}. Also, if Fe does not occupy interstitial position, it will not migrate out of MgO to form metallic precipitate as shown in some experiments \cite{Wuensch_1962,Molholt_2014}. 

\section{Conclusions}
In this work, we have carried out a systematic investigation of stability of implanted TM dopants in MgO and BaO, using density functional theory, for a dilute doping limit. We have calculated the TM dopant formation energies in various charge states as a function of the electronic chemical potential. We show that TM dopants can occupy interstitial site, in contrary to the common belief that TM dopants will invariably substitute cations in the oxide. In case of BaO, all the TM dopants prefer interstitial site. However, for MgO, Fe, Co, Ni, Cu and Zn prefer interstitial sites while Sc, Ti, V, Cr and Mn prefer to substitute Mg atoms. Our results suggest that the site preference of TM atoms in rock-salt oxides depends on relative ionic radii and oxygen affinity of the host cation and the dopant. If ionic radius of the host cation is significantly bigger than TM atom, then the dopants prefer interstitial sites. However, if ionic radius of the dopant is comparable to that of the host cation, it can substitute lattice cations only if its oxygen affinity is similar to that of the host cation. Stability of dopants in oxides depends on the electronic chemical potential; lower electronic chemical potential leads to higher stability. For given electronic chemical potential, higher the affinity of the TM to oxygen, greater the stability of the dopant. Our result on Ni occupying interstitial site explains experimentally observed phenomena of implanted Ni ions migrating out of MgO after annealing. Experimentally reported charge states of implanted Fe in MgO was found to be stable only when Fe is in interstitial site. Similarly, various experiments on ions implanted in stable oxides, for example Al$_2O_3$,~\cite{Xiang2004,Stepanov2005,ALVES200355,MCHARGUE19981,MOTASANTIAGO2012574} can be explained in the light of our findings using the scheme we followed to explore preferred defect sites in the host lattice. Computational route presented here can be applied to other stable oxides like CaO, ZnO and various perovskites to explore the possibilities of stabilizing dopants at interstitial sites. Stabilizing defects at interstitial sites can give rise to various exciting phenomena, for example unusual exchange coupling between dopants at interstitial sites. TM dopants at interstitial site in stable oxide like MgO can transfer charge to single atom Au and Pt, anchored on the surface. This could lead to better catalytic activity of single atom Au and Pt bound to such stable oxides. Thus this study warrants design of experiments so that implanted ions can be stabilized in interstitial sites.  
 
\begin{acknowledgements}
 D.M. gratefully acknowledges the support of post-doctoral fellowship provided by IIT Madras, India. We acknowledge Dr. Somnath Bhattacharya's help in providing us access to VASP source code. The authors acknowledge helpful discussions with Dr. Blas P. Uberuaga and are grateful for his valuable suggestions. We further acknowledge the Computer Center, IIT Madras, for providing computational facilities.
\end{acknowledgements}

\bibliographystyle{ieeetr}
\bibliography{defect}

\end{document}